\documentclass[a4paper,11pt]{article}
\pdfoutput=1 
\usepackage{jheppub} 

\usepackage[T1]{fontenc} 
\usepackage{amsmath,amssymb,euscript}
\usepackage{braket}
\usepackage{slashed}
\usepackage{xspace}
\usepackage{color}
\usepackage{hyperref}
\usepackage{epsfig}
\usepackage{xcolor}
\usepackage{xspace}
\usepackage{verbatim}
\usepackage{multirow}
\usepackage{booktabs,graphicx}
\usepackage{mathtools}
\usepackage{tabulary}
\usepackage{soul}
\usepackage{caption}
\usepackage{subcaption}
\usepackage{orcidlink}
\usepackage{xfrac}
\usepackage{float}
\usepackage{placeins}
\usepackage{enumitem}
\usepackage{amssymb}

\arxivnumber{2406.08557}
\preprint{DESY-24-090}

\title{\boldmath SHIFT@LHC: Searches for New Physics with Shifted Interaction on a Fixed Target at the Large Hadron Collider}

\author[a]{Jeremi Niedziela \orcidlink{0000-0002-9514-0799}} 

\affiliation[a]{Deutsches Elektronen-Synchrotron DESY, Notkestr. 85, 22607 Hamburg, Germany} 

\emailAdd{jeremi.niedziela@cern.ch}

\abstract{New low-mass particles with very small couplings to standard model particles that travel significant distances before decaying are interesting candidates to address some of the most intriguing questions of modern physics. In this paper, I propose to extend the LHC's research program by installing a gaseous fixed target referred to as SHIFT at around 160~meters from the CMS interaction point. When the LHC proton beam collides with this target, interactions at a center of mass energy of $\approx$113~GeV would occur. The particles produced in such collisions, or their decay products, would travel through the rock and other material on their path, potentially reaching the CMS detector where they can be registered and studied. Such an approach would allow us to access otherwise uncovered regions of~parameters phase space at a relatively low cost since it does not require constructing a new detector. Various aspects such as angular and lifetime coverage or material survival probability have been studied. The results are interpreted within two new physics models, namely, the Dark Photons and the Hidden Valley scenarios, and compared with the standard proton-proton physics program of CMS. A comparison is also made with the fixed target program at LHCb, as well as parasitic detectors such as FASER or \mbox{MATHUSLA}. The obtained results indicate that, despite assuming just 1\% of the nominal CMS lu\-mi\-no\-si\-ty to be available to SHIFT, the physics reach could be extended by a factor of up to 150 (1000) for Dark Photon (Hidden Valley) scenarios, depending on the signal model parameters. }

\begin{document} 
\maketitle
\flushbottom

\section{Introduction}
\label{sec:intro}

The standard model (SM) of particle physics describes the fundamental building blocks of matter and the interactions between them. It has been experimentally tested and has proven to correctly predict the outcome countless times. However, despite its success, the SM is widely believed to be incomplete. The Large Hadron Collider (LHC) is an extremely versatile tool, allowing to search for a very wide range of particles and phenomena that could shed light on some of the mysteries of modern physics, such as the nature of Dark Matter, matter-antimatter asymmetry of the Universe, and the neutrino masses, to name a few. Historically, such searches have been focused on high-mass (multi-TeV) particles. However, in light of the lack of new physics discoveries since the discovery of the Higgs boson in 2012, the High-Energy Physics community started to shift, or expand, towards particles with lower masses and extremely tiny couplings, or unusual signatures, which could have escaped the more traditional searches.
    
One of the interesting regions to search for new physics is long-lived particles - in this paper we will use this term for particles that travel more than $O(100\mu m)$ after being produced at the interaction point, and prompt otherwise. Given there are plenty of long-lived particles in the SM, there is no reason to assume that beyond the SM (BSM) particles should decay promptly. Such particles could create interesting signatures in the detector, including disappearing tracks, displaced vertices, delayed jets or photons, and many more \cite{Alimena_2020, LEE2019210}. The interest in those signatures has been growing in the past years, with many LHC experiments both large (ATLAS, CMS, LHCb \cite{atlas, cms, lhcb}) and small (e.g. MoEDAL, FASER, SND, milliQan \cite{Pinfold:1181486, fasercollaboration2022faser, lhccollaboration2023sndlhc, Haas_2015}) joining the effort, and more being proposed or under construction (such as CODEX-b, MATHUSLA, ANUBIS, or FACET \cite{Aielli_2020, Chou_2017, bauer2019anubis, Cerci:2021nlb}).

Another direction that has started to be investigated more in the past years is particles with relatively low masses (up to the GeV scale) that are very weekly coupled to the SM, making them difficult to observe in the large LHC detectors. Unlike the TeV-scale particles, these low-mass ones would be produced mostly in the forward direction, which is not particularly well covered by the large experiments. As a result, more and more experiments aiming at the forward region are being proposed, such as the aforementioned FASER and SND already producing physics results, or a whole new underground cavern being proposed for the Forward Physics Facility (FPF) \cite{Feng_2023}, which would host a few experiments focusing on searches for new physics in this region.

In this paper I propose an experiment aiming at a combination of these two interesting regions, searching for new long-lived particles in the 11-60 GeV mass range, produced in the forward direction. I propose to install a fixed target at the LHC at some distance (at the order of 100~meters) from one of the existing large LHC detectors. The project is referred to as SHIFT@LHC, for Shifted Interaction on a Fixed Target at the Large Hadron Collider. As will be discussed in Sec.~\ref{sec:experimental_setup}, the installation of a fixed target at the LHC has already been considered by ALICE and LHCb and was successfully deployed by the latter (called SMOG/SMOG 2), proving that such a project is feasible \cite{Hadjidakis_2021, Bursche:2649878, CERN-LHCC-2019-005, Franzoso:2825146, Massacrier_2018}. What is also worth noting is the cost: the estimate for SMOG 2 is at the level of 200 kCHF, which can be compared to over 1 MCHF for FASER \cite{fasercollaboration2018technical}, $>$10 MCHF for ANUBIS \cite{bauer2019anubis}, and as much as 40 MCHF for the Forward Physics Facility \cite{ANCHORDOQUI20221}. Since SHIFT only requires a fixed target, but no new detector is needed, it would be similar to SMOG 2 from the technical point of view - one would expect a cost at the level of a few hundred kCHF, which is relatively inexpensive compared to other proposals.

We will use the CMS detector as an example, however in principle, any of the large LHC detectors (ATLAS, ALICE, or LHCb) could be used, with the physics reach changing depending on the luminosity and detector's characteristics. The collisions happening at SHIFT result in the production of particles, traveling roughly in the direction of the detector, which can be stopped in the rock or other material on their path, decay in flight, or reach the detector and have a chance of being registered. In this work, we will focus on new physics scenarios with pairs of muons in the final state, which have a high chance of going through the material unaffected and for which CMS has excellent detection capabilities (and large angular coverage). Given the fixed distance between the target and the detector, the time of arrival of the products of the interaction can be exploited for triggering and possible beam-halo, beam-gas, and cosmic background suppression. 

It is also important to emphasize that any other final state could also be studied with SHIFT, for instance, those containing photons, electrons, and jets in the final state. Especially interesting to consider is that a fixed target would naturally contain not only protons but also electrons, which would allow for quark-electron collisions, greatly amplifying cross-sections for direct production of leptoquarks. These alternative final states however come with several challenges: more precise energy loss studies would be necessary and the overall rock-survival probability would be much lower than for muons; events could be similar to radioactive nuclei decays in the cavern walls, requiring dedicated studies and careful background rejection; the cross-section of detectors able to register e.g. photons is generally much smaller than this of the muon detectors. Although in principle all these challenges could be overcome, in this paper we will focus on the simplest scenario, i.e. processes with muons in the final state.

Aspects such as the angular and lifetime acceptance, interaction with the rock, but also integration with the main LHC physics program, will be discussed. The physics potential will be studied using two BSM physics scenarios: Dark Photons and Hidden Valley models. To put the discovery reach in some perspective, we will focus on a scenario where SHIFT is installed $O(100~\rm{m})$ downstream of the CMS interaction point, and compare the results to what could be achieved at CMS with the default collider-mode proton-proton program, under the same assumptions and simplifications. I will show that such an experiment allows one to access uncovered regions of BSM parameters phase-space, without the need to build new detectors or experimental caverns, therefore providing a~relatively inexpensive way of extending the LHC's research program.

Experiments such as FASER or MATHUSLA have also been considered, however with their specific acceptance, using the collider mode of the LHC, and other characteristics they are more suited to search for particles with masses up to $\approx$1~GeV, which is an order of magnitude below the focus of this study, and therefore will not be mentioned further in this paper. The existing fixed target experiments at the LHC: SMOG and SMOG 2 will be discussed, although as it will be explained, they are less sensitive to the studied new physics scenarios than the proposed SHIFT project.

\begin{figure}[ht]
\centering
\includegraphics[width=.7\textwidth, trim={0 0 0 0}, clip]{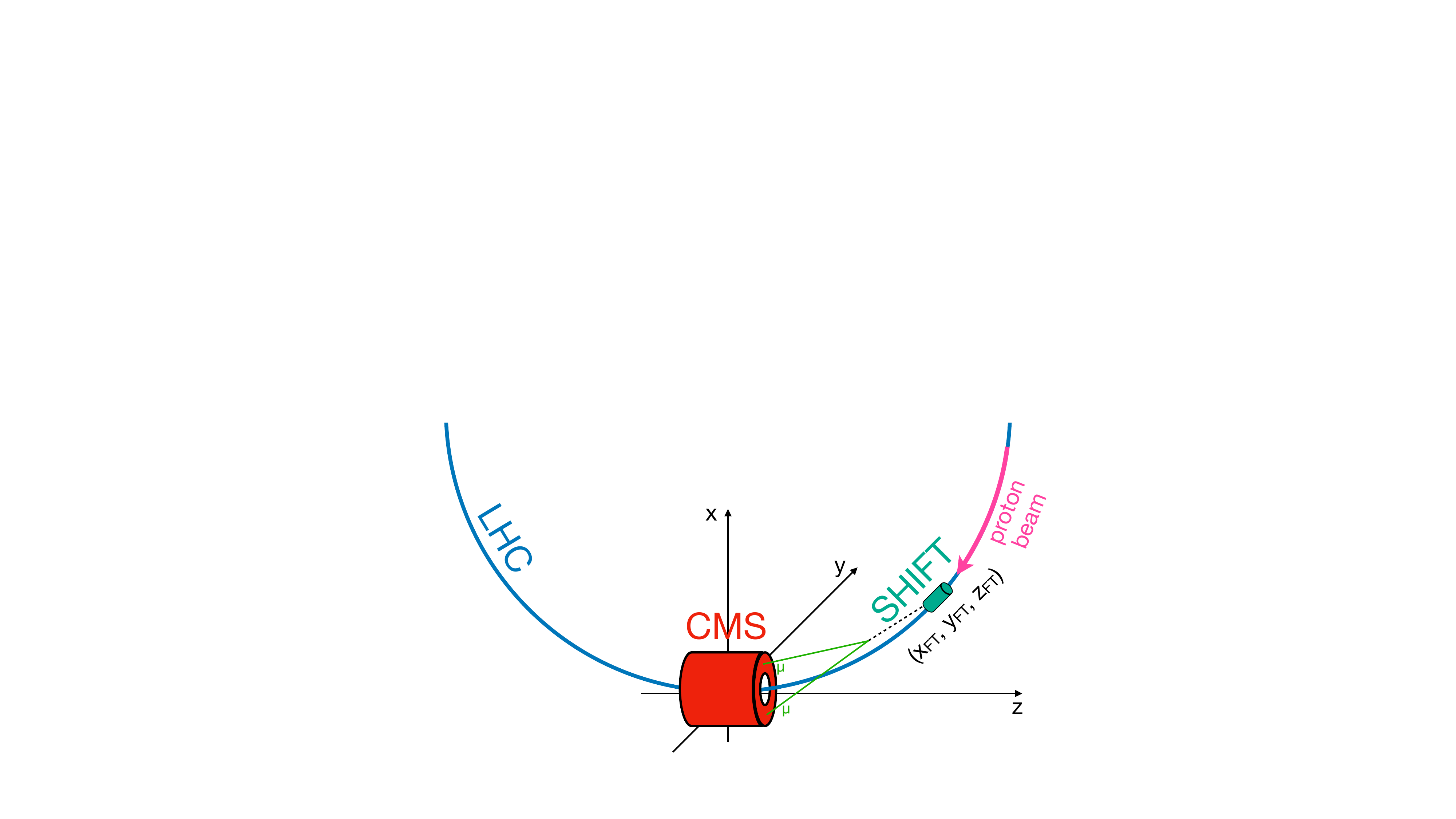}
\caption{\label{fig:layout} The general layout of SHIFT w.r.t. the LHC ring and the CMS detector.}
\end{figure}

\clearpage
\newpage

\section{Experimental setup and assumptions}
\label{sec:experimental_setup}

In this section, the details of the experimental setup will be discussed, such as the assumptions on the fixed target design, its location along the LHC ring, and the expected integrated luminosity.

\subsection{Fixed target and Luminosity}

The idea of placing a fixed target at the LHC is by no means new: both ALICE and LHCb considered it, with the latter actually designing, building, and successfully operating one (although more in the context of heavy-ion, hadron, and spin studies) \cite{Hadjidakis_2021, Bursche:2649878, CERN-LHCC-2019-005, Franzoso:2825146, Massacrier_2018}. SMOG and its successor, SMOG 2, are essentially chambers located at the LHCb interaction point that can be filled with different gases, allowing to control the density, and therefore also the rate of collisions. In this work, we will build on the knowledge gained by SMOG, assuming that a similar fixed target could be placed near the CMS (or any other) interaction point. However, unlike SMOG, we place it at a distance of the order of 100~meters from the collider interaction point, which is optimized to position the detector in the peak of the rapidity of particles produced in the collision, and has an additional advantage of the rock and other materials acting as a shield, filtering out some of the background (especially QCD, as shown in Sec.~\ref{sec:selection_efficiency}).

The energy of the LHC proton beam is assumed to be 6800 GeV, colliding with a~stationary target. To simplify the generation of signal and background samples, and make the study more comparable to the standard proton-proton LHC physics program, we will use a target made of protons (similar to the SMOG's hydrogen variant, since the electrons are irrelevant for the studied physics scenarios). We will also assume that the gas density can be tuned to achieve the desired total integrated luminosity. Installation of a fixed target at the LHC shall not significantly disturb the main physics program - in the case of SMOG 2 the collected luminosity was below 5\% of the main LHCb's program \cite{Bursche:2649878}, therefore we will conservatively assume that SHIFT can collect 1\% of the nominal CMS luminosity. With the CMS integrated luminosity in Run 4 expected to reach $715~fb^{-1}$ \cite{lumi_run4}, we will assume an integrated luminosity of $7.15~fb^{-1}$ for SHIFT. This number would be similar for ATLAS but would have to be significantly reduced if one wanted to use LHCb or ALICE, which typically reach total integrated luminosity at least an order of magnitude lower than ATLAS and CMS. What is also worth stressing is that despite the comparably large luminosity assumed for SHIFT, the corresponding relative beam loss is still at the level of $\approx$0.4\% over a 10-hour long run, assuring that the main physics program remains virtually unaffected.

The general layout of SHIFT w.r.t. to the LHC ring and the CMS detector is presented in Fig.~\ref{fig:layout}. The exact location of the target can be tuned, but in general, a distance of around 100-200 meters allows to maximize the number of particles produced in such interactions to reach CMS - due to the asymmetric system, most particles are produced at forward rapidity of $\eta\approx5$, from which a reasonable distance can be estimated. In reality, the location would be highly limited by the machine's instrumentation, and therefore a detailed study of possible installation points would be necessary. In this paper, the location of the target is described by the distance along the CMS's z-axis, and required to be along the LHC ring (which is assumed to be a perfect circle). Distances between 30 and 1000 meters were considered, and 160 meters was found to be a good choice for a wide range of physics scenarios (see Sec.~\ref{sec:limits} for more details). This location also avoids interference with the beam pickup monitors (BPTX), the CMS-TOTEM Precision Proton Spectrometer (CT-PPS) detectors, and the superconducting dipole magnets, which are all located at and beyond a distance of 175 meters - the products of the collision with the fixed target would move away from those and towards the CMS detector. The interference with the HL-LHC equipment in front of the target is also minimal - the nearest magnet (D2) is located 6 meters away, and given its size and the bulk of collision products produced at $\eta\approx5$, most particles will miss this magnet by $\approx2.5$~cm. It is also worth emphasizing that the location of the target can be varied to some extent (e.g. due to the machine's restrictions), still preserving good acceptance for certain BSM scenarios. In reality, the LHC tunnel is straight up until $\approx$230 meters from the interaction point, which results in a smaller amount of rock between the target and the detector, and slightly different angle between the target and the longitudinal axis of the detector. On the other hand, there is other material, including magnets, support structures, but also absorbers, which should be considered in a full simulation of the setup. As will be explained in Sec.~\ref{sec:selection_efficiency}, the rock does not play a significant role for muons (at distances of the order of a few hundred meters), while just 1~meter of rock would completely stop hadrons, and therefore the uncertainty in the survival probability has a marginal impact on the results. Similarly, it has been verified that small changes in the target location and orientation do not change the acceptance in a significant way. Finally, one should notice that slight decrease in sensitivity for the studied BSM scenarios due to the exact placement of the target may result in an increased sensitivity for other signal models, for which rapidity distribution or lifetime may differ.

\subsection{Signal and background samples}

As explained in Sec.~\ref{sec:intro}, we will focus on BSM signal models with new particles decaying into pairs of muons, which can travel through hundreds of meters (depending on the energy) of rock and other obstacles, making them suitable for this kind of search. The Monte Carlo samples are generated with \verb|PYTHIA8| \cite{Sj_strand_2015} for both CMS (collider mode with center of mass energy of 13.6 TeV) and SHIFT (fixed target mode with a 6.8 TeV beam, corresponding to $\approx$113 GeV center of mass energy). The following dominant backgrounds will be considered:

\begin{itemize}[noitemsep, topsep=0pt, label=$\diamond$]
    \item Quantum Chromodynamics (QCD): gluon/quark mediated processes, typically followed by hadronization and subsequent hadron decays, which may result in muons in the final state. Typical settings were used for this background, with samples produced in bins of the hard process transverse momentum $\hat{p_{T}}$.
    \item Drell-Yan (DY): processes mediated by on/off-shell Z-boson or a virtual photon, with two muons in the final state.
\end{itemize}

\noindent In principle, one should consider two other potential sources of background: cosmic muons and beam-halo/beam-gas collisions. The cosmic background can be easily suppressed requiring that the origin of the incoming muon is roughly consistent with the LHC's plane - such horizontal cosmic muons would have to traverse many kilometers of rock, and the chance for two muons doing so at the same time and forming a common vertex is negligible. The beam-halo and beam-gas is much more interesting though: protons in the LHC beam may collide with collimators or the residual gas in the accelerator's pipe, creating the same signature as SHIFT, although with collision points located anywhere along the LHC, rather than accumulated on the fixed target's location. On the one hand, such events can be suppressed by requiring that the reconstructed mother of the dimuon is consistent with originating from SHIFT. On the other hand, the beam-halo/beam-gas events can be very useful for the development of the trigger, reconstruction, and analysis techniques. In this study, we will neglect the cosmic and beam-halo backgrounds.

\begin{figure}[ht]
\centering
\includegraphics[width=.99\textwidth, trim={0 0 0 0}, clip]{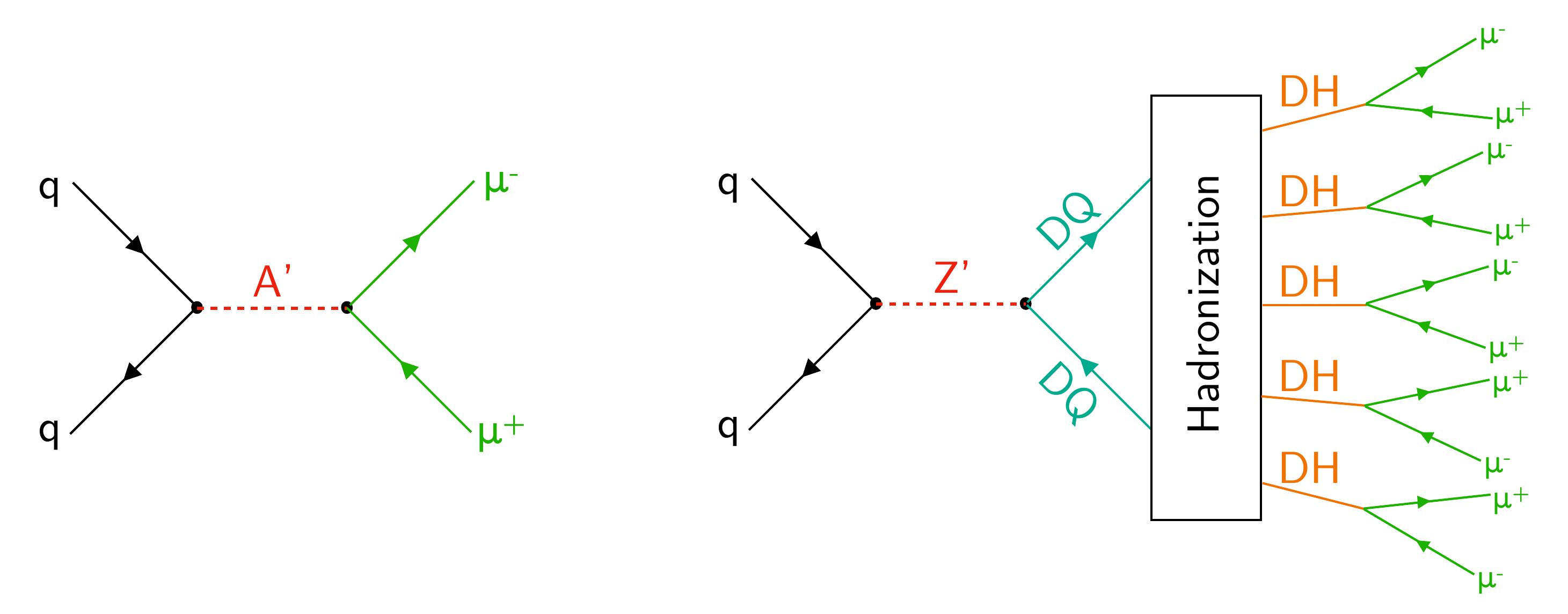}
\caption{\label{fig:diagrams} Diagrams representing the two studied BSM scenarios. Left: Dark Photon A' produced in a proton-proton collision, decaying to a pair of muons. Right: a Z' boson produced in proton-proton collisions, decaying to a pair of Dark Quarks that undergo hadronization within the Dark Sector, resulting in several Dark Hadrons, subsequently decaying to pairs of muons.}
\end{figure}

\noindent For signal, two different scenarios are studied (as depicted in Fig.~\ref{fig:diagrams}):

\begin{itemize}[noitemsep, topsep=0pt, label=$\diamond$]
    \item Dark Photons (DP): a model provided by \verb|PYTHIA8| introducing a new gauge boson A' which couples directly to SM particles \cite{Sj_strand_2015}. The default settings of the \verb|NewGaugeBoson| class are used, except for vector and axial couplings to muons tuned to better reproduce the SM invariant mass distribution slope: $v_{\mu} = -0.04$, $a_{\mu}=-0.5$, and couplings to SM quarks set to $v_{d} = a_{d} = -0.04$ and $v_{u} = a_{u} = 0.02$. Decays to muons are forced in 100\% of events. A few benchmark points were chosen to demonstrate differences in kinematics: $m_{A'}\in \{5, 30\}$ GeV and $c\tau \in \{10^{-3}, 10^{1}, 10^{3}\}$ m, with more mass and mean proper decay length values used for limits setting.
    \item Hidden Valley (HV): a more complex model introducing an equivalent of the strong interaction in the Dark Sector, with a new gauge boson Z', dark quarks DQ, and dark hadrons DH \cite{Carloni_2010, Carloni_2011}. A relatively simple scenario is considered, in which SM fermions fuse to create the Z' boson, which decays to two DQs hadronizing in the dark sector into some number of DHs, each decaying back to a pair of SM muons. Two benchmark points were selected with $m_{Z'} = 60$~GeV, $m_{DH} \in \{5, 20\}$~GeV, $m_{DQ} = 1$~GeV, $c\tau=10^{-1}$~m, and more mass and mean proper decay length points were used to set limits.
\end{itemize}

\noindent For all backgrounds and signals the default \verb|PYTHIA|'s constraints on lifetime were removed completely (for SHIFT) or extended to the detector's boundary (for CMS) to allow for very long-lived particles to still decay tens or hundreds of meters away from the production point. The mean proper lifetime is then set manually for all considered BSM particles decaying to muons. Depending on the scenario and the physics process, between 100K and 100M events per sample have been produced.

\begin{figure}[t!]
\centering
\includegraphics[width=.49\textwidth, trim={0 0 70pt 0}, clip]{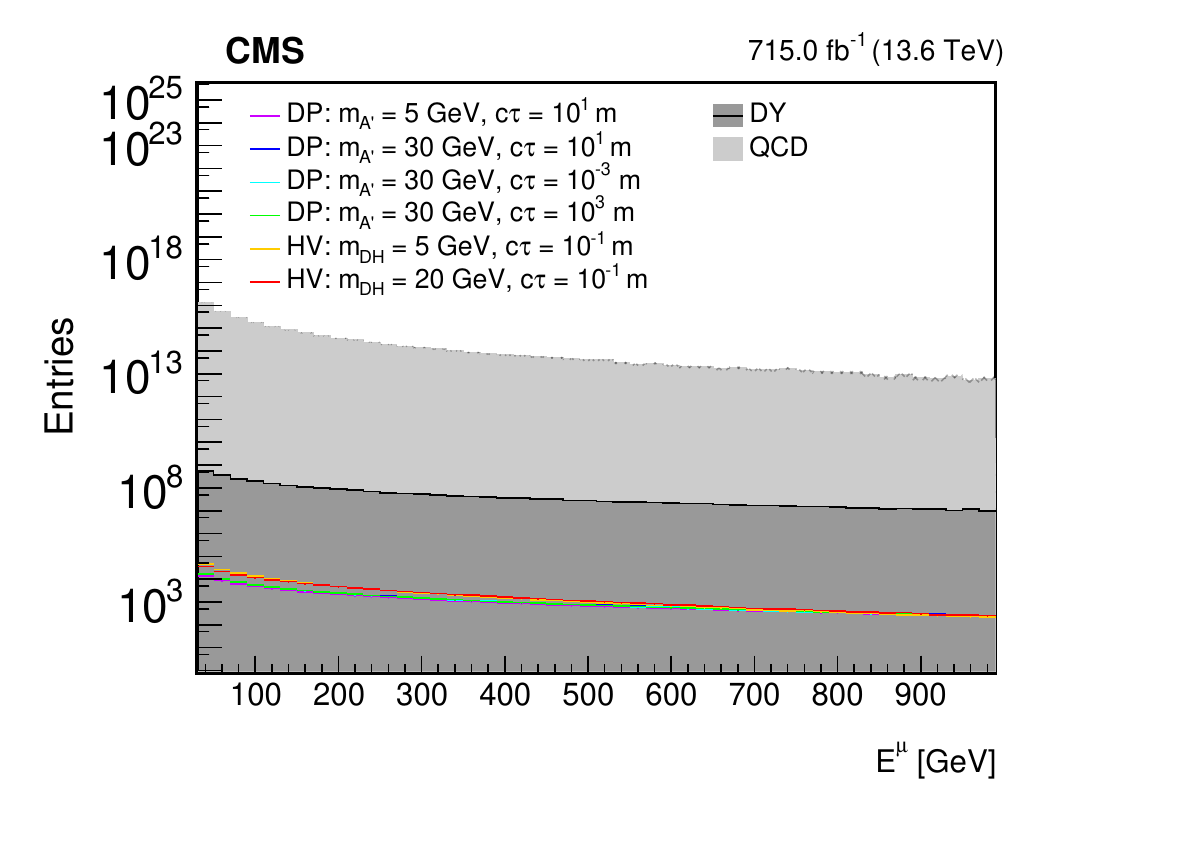}
\includegraphics[width=.49\textwidth, trim={0 0 70pt 0}, clip]{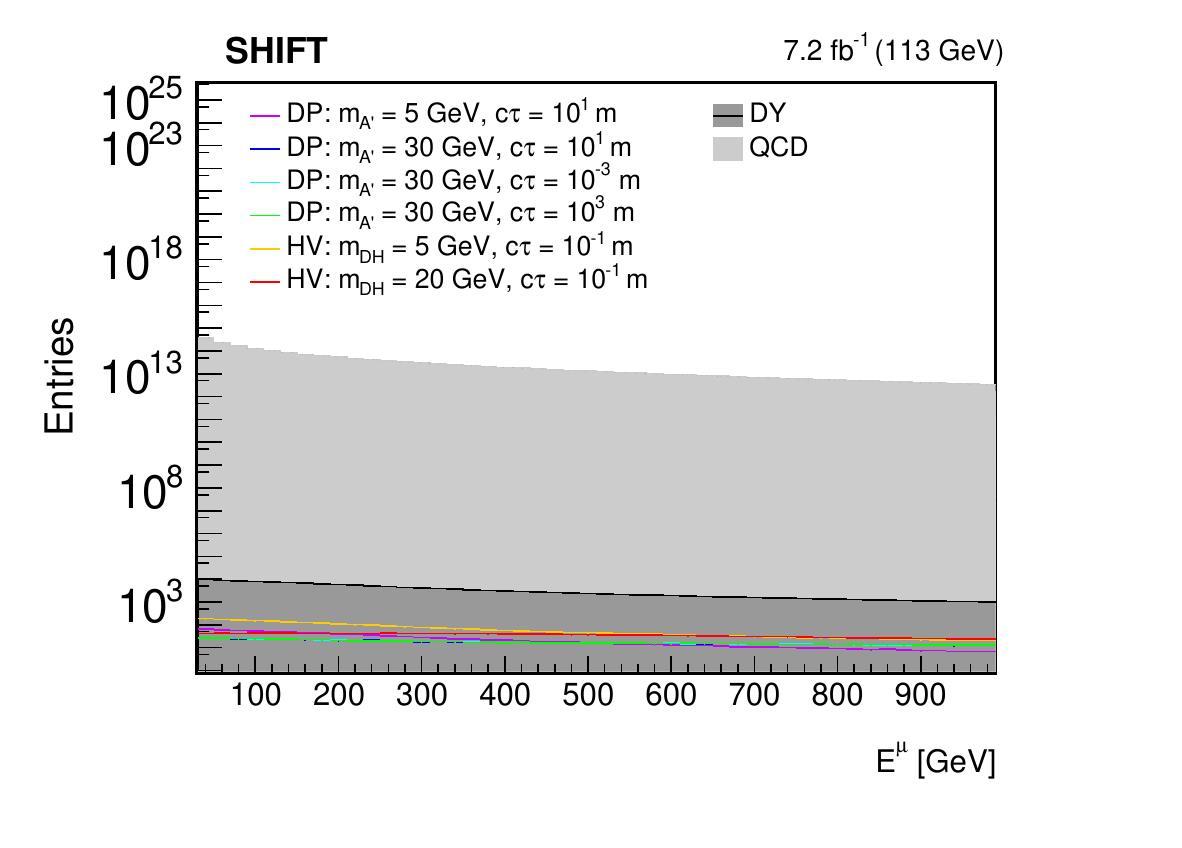}
\includegraphics[width=.49\textwidth, trim={0 0 70pt 0}, clip]{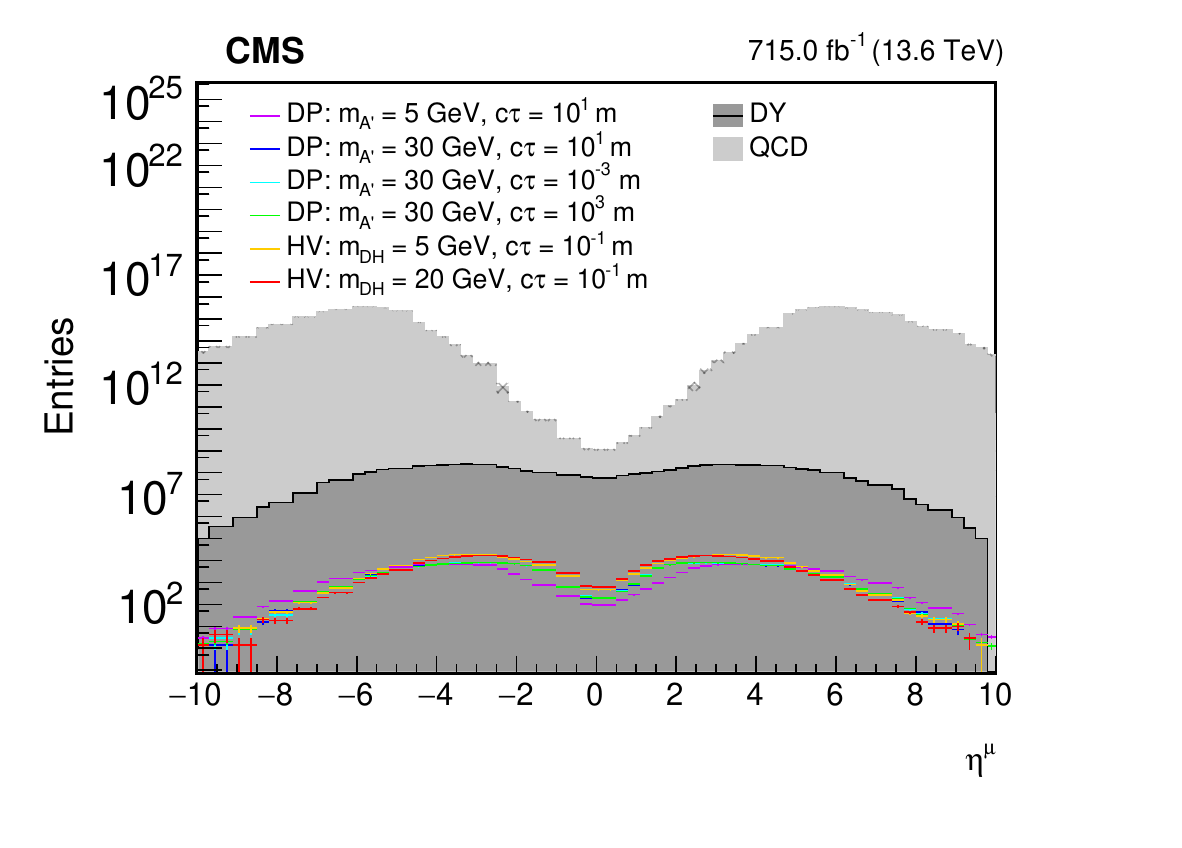}
\includegraphics[width=.49\textwidth, trim={0 0 70pt 0}, clip]{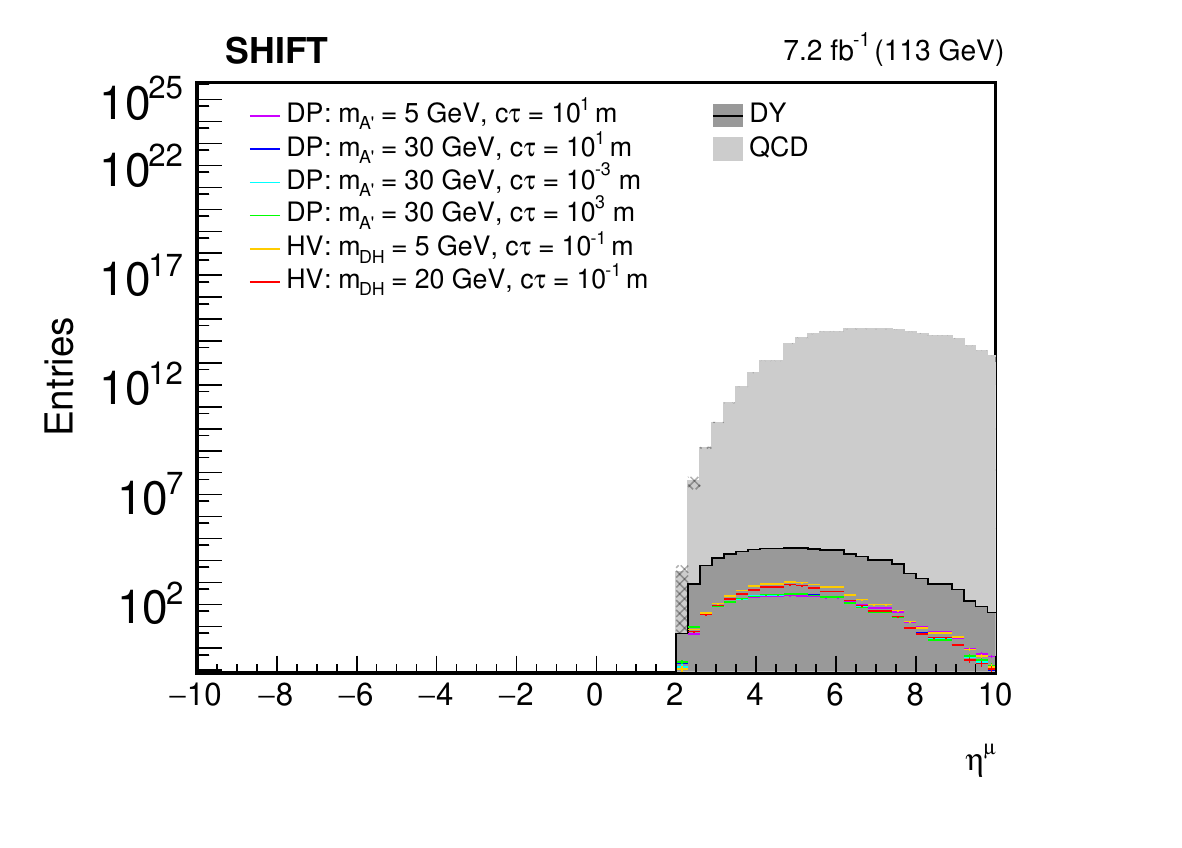}
\caption{\label{fig:initial_kinematics} Single-muon kinematics before event selections. Left: CMS scenario, right: SHIFT scenario. All muons are required to have total energy above 30 GeV. Distributions are normalized to cross-section and luminosity ($715~\rm{fb}^{-1}$ for CMS and $7.15~\rm{fb}^{-1}$ for SHIFT), with cross-section set for all signals to $0.2~\rm{pb}$~(nb) for SHIFT (CMS) for visual purposes.}
\end{figure}

\begin{figure}[t!]
\centering
\includegraphics[width=.49\textwidth, trim={0 0 70pt 0}, clip]{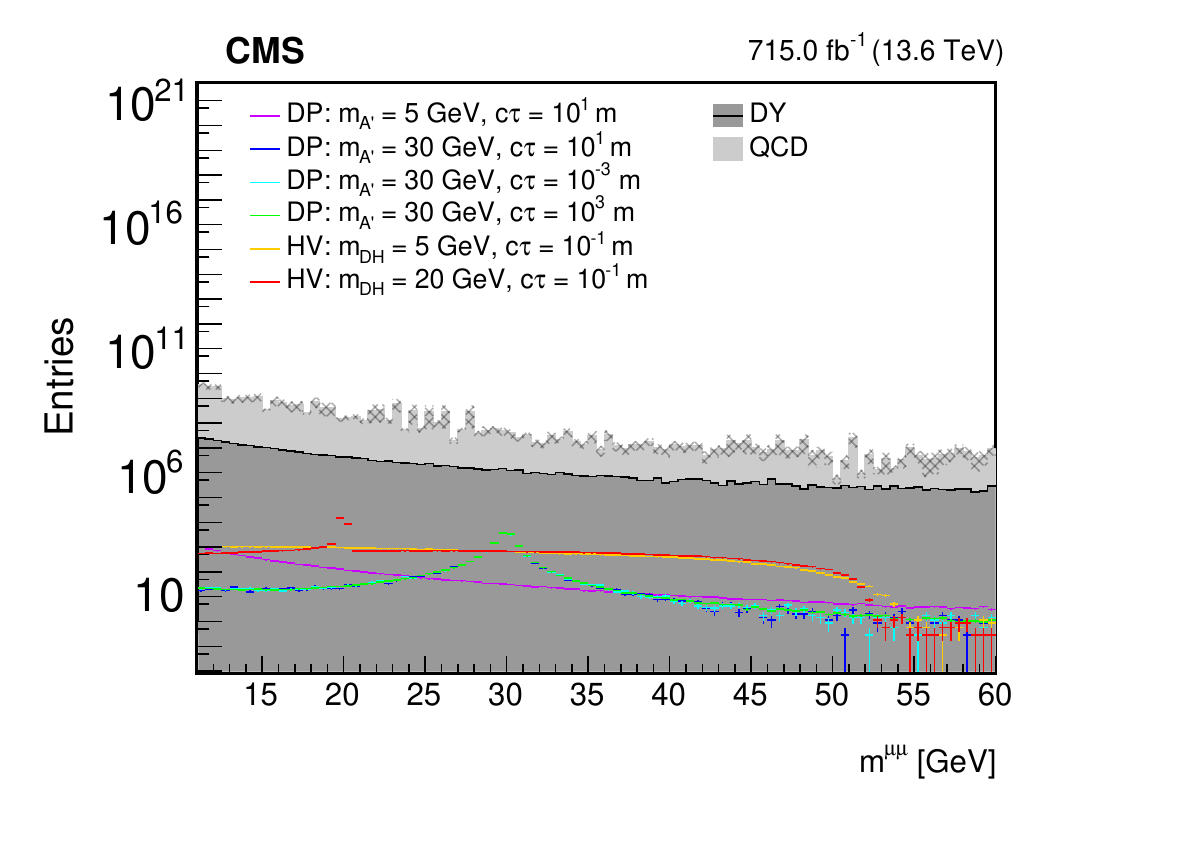}
\includegraphics[width=.49\textwidth, trim={0 0 70pt 0}, clip]{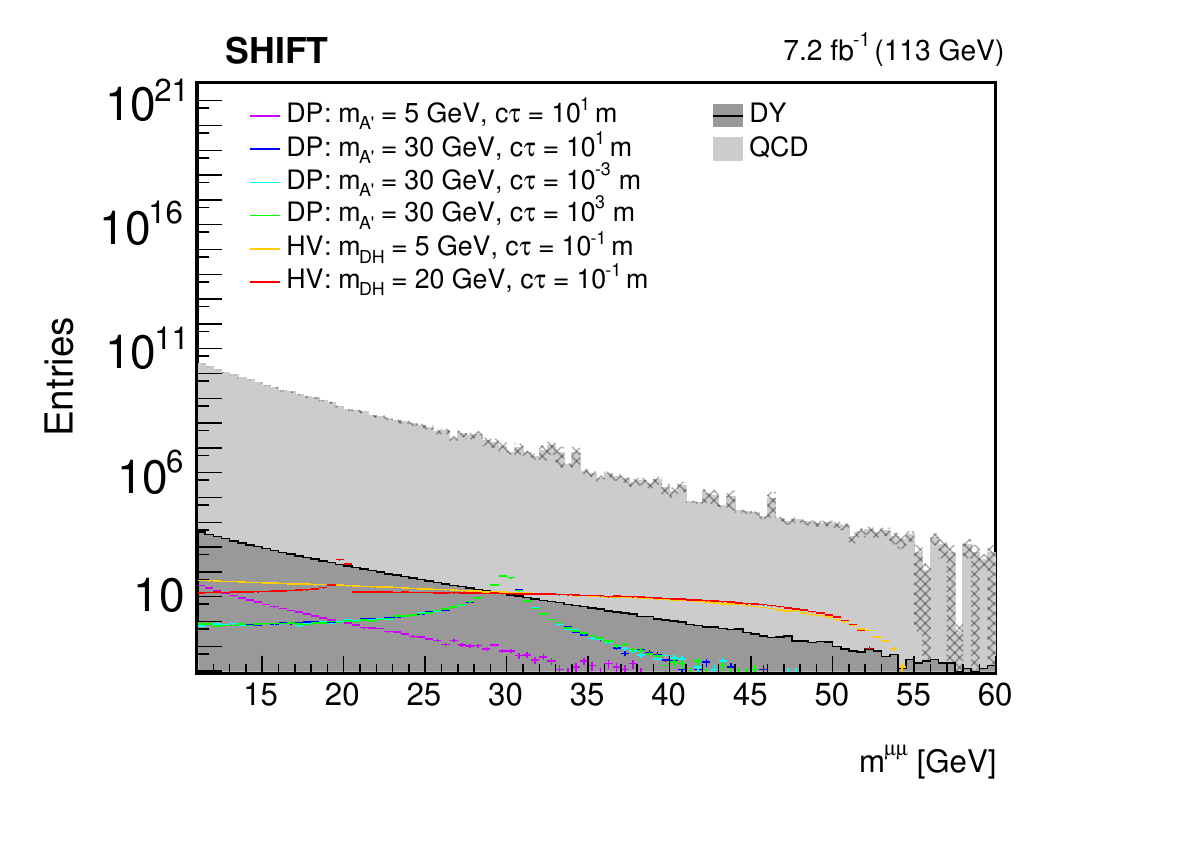}
\caption{\label{fig:initial_pair_kinematics} Dimuon invariant mass before event selections. Left: CMS scenario, right: SHIFT scenario. All muons are required to have total energy above 30 GeV and dimuon invariant mass above 11 GeV. Distributions are normalized to cross-section and luminosity ($715~\rm{fb}^{-1}$ for CMS and $7.15~\rm{fb}^{-1}$ for SHIFT), with cross-section set for all signals to $0.2~\rm{pb}$~(nb) for SHIFT (CMS) for visual purposes.}
\end{figure}

As will be explained later in Sec.~\ref{sec:selections}, all muons are required to have energy $E^{\mu} > 30$~GeV and dimuons to have the invariant mass $m^{\mu\mu} \in [11, 60]$~GeV. The kinematic distributions for muons/dimuons with these requirements met (but without any other selections) for both CMS and SHIFT are shown in Fig.~\ref{fig:initial_kinematics} for single muons and Fig.~\ref{fig:initial_pair_kinematics} for dimuons. As expected, in the fixed target scenario of SHIFT, the pseudorapidity distributions are shifted towards higher values for both background and signal. It is worth noting that in the CMS scenario muon $\eta$ peaks at larger values between 2 and 6, which is caused by the muon energy requirement of 30 GeV. The difference in available energy between the collider and fixed-target modes can be very clearly seen in the invariant mass, where the distribution for SHIFT falls much quicker than that for CMS.

\section{Event selections}
\label{sec:selections}

The selection of events in this work focuses on the chance for dimuons to reach the detector, given the angular and lifetime acceptance, as well as interactions with the rock and other obstacles. No sophisticated selections specifically aimed at background reduction are applied. Depending on a specific physics scenario, one could exploit features such as the location of the reconstructed dimuon vertex, muon energy, number of muon pairs, etc., which, however, is not the focus of this study. Instead, all events passing acceptance and trigger/reconstruction criteria will be considered and a bump hunt in the invariant mass spectrum of surviving dimuons will be used for limits setting.

\subsection{Angular and lifetime acceptance}
\label{sec:acceptance}

One of the main considerations for experiments with detectors located at a distance from the interaction point is the solid angle coverage. The further we place the target from the detector, the lower the acceptance. What does play in SHIFT's favor though is the collision asymmetry, resulting in most particles traveling in the general direction of the detector. Another important factor is the angular distance between the two muons - they both have to reach the detector, therefore small $\Delta\eta^{\mu\mu}$ is desired. This depends on the model details, but in general, for the considered scenarios $\Delta\eta$ is small, while $\Delta\phi$ peaks at $\pi$ (which means that the two muons will likely hit two opposite halves of the detector).

The procedure to check whether a given dimuon intersects with the detector is the following. The detector is modeled as a cylinder with a cylindrical opening in the middle (corresponding to the $\eta$ coverage of the detector), characterized by the outer radius $R_{D}^{O}$, the inner radius $R_{D}^{I}$, and the total length $l_{D}$. For CMS these parameters are assumed to have values of $R_{D}^{O} = 7.5$~m, $R_{D}^{I}$ is determined from $\eta_{max}^{CMS} = 2.4$, and $l_{D} = 22$~m. The~location of the fixed target is described by the distance along the z-axis of CMS, $z_{FT}$, with the x-coordinate $x_{FT}$ calculated such that it is along the LHC ring (modeled as a perfect circle with a radius of 4.3 km), the y-coordinate unchanged, i.e. $y_{FT} = 0$~m, and rotated to be tangent to the ring (see Fig.~\ref{fig:layout} for a graphical illustration). All particles generated by \verb|PYTHIA8| undergo translation and rotation such that the CMS detector is placed at (0,~0,~0) and the collision point is at ($x_{FT}$, $y_{FT}$, $z_{FT}$), and tangent to the LHC ring. Finally, for each particle, intersections with the inner/outer side or any of the end-caps of the cylinder are verified - the particle passes the angular acceptance if such an intersection is found. 

This is a simplified picture, neglecting for instance the length of the trajectory inside of the detector, and a proper simulation (even simplified) would be necessary for a more precise determination of whether such muon could be reconstructed. However, it would be a relatively small effect, especially for SHIFT, where most particles travel almost parallel to the detector's z-axis. It would have a larger impact on the CMS scenario, in which particles close to the $\eta$ range limit would intersect with the cylinder, but could only cross one or two layers of detectors, making it impossible to reconstruct them.

Another aspect to consider is the lifetime acceptance. For the CMS scenario, the decay must happen within the detector, limiting the maximum decay length to a few meters. An~important advantage of SHIFT is that the decay may occur anywhere between the fixed target and the end of the detector, potentially increasing the maximum allowed decay length to hundreds of meters. The only requirement related to lifetime acceptance imposed on the events is that the dimuon vertex is not further than 2 meters from the detector's center (assuring enough muon detector layers to still be well reconstructable). This criterion is applied together with the angular acceptance explained above and they will together be referred to as "acceptance".

\subsection{Muon interactions with the surroundings}

For muons produced at the fixed target to reach CMS, they must survive traversing tens or hundreds of meters of rock and other material. Though muons' interactions are relatively weak, they cannot be completely neglected. A simple study was performed using \verb|GEANT4| \cite{GEANT4:2002zbu} simulation of different volumes of standard rock, confirming the findings of \cite{Yuan_2022} - the energy loss for muons is almost negligible until they rapidly lose all their energy after traveling some characteristic distance $d_{crit}^{\mu}$ in the rock. Since in reality some fraction of the muon's path would cross the LHC magnets, cryostats, cables, and support structures, but also the air in the tunnel, concrete walls, etc., a dedicated study would be needed to fully simulate the expected energy loss and survival probability. To keep things simple for this work, we will assume that muons traverse the rock unaffected as long as they have enough energy not to be stopped completely. This critical energy is proportional to the distance they have to travel: $E_{crit}^\mu = a\cdot d_{crit}^\mu + b$, and here, based on the results presented in \cite{Yuan_2022}, we assume $a = 0.5$~GeV/m and $b = 1$~GeV. Since I propose to place the target at around 160 meters, muons with energy above around 80~GeV should reach the detector, which corresponds to the vast majority of them, and therefore has a small effect on the selection efficiency. 

One should also consider that most background muons come from pions, which have lifetimes large enough that they can travel tens of meters before decaying. It was verified with the same \verb|GEANT4| simulation that even the most energetic pions (and other hadrons) could not traverse more than $\approx1$~meter of standard rock. If such a pion is stopped in the material, the resulting muon will be emitted in a random direction and with very low energy, therefore itself being quickly absorbed. This effect has also been included in this analysis.

Finally, we will neglect the effect of the LHC's magnetic field on these muons, given that the very strong field inside of the dipoles decays very rapidly with increasing distance from the beam axis, and the muons and other particles produced at the fixed target are not expected to exactly follow the beam's direction. Nevertheless, the effect of the LHC magnets should be studied in detail for a more precise estimation. The requirement of having enough energy to pass through the material between the collision point and the detector will be referred to as "material survival".

\subsection{Triggering and reconstruction}

Muons reaching the CMS detector also need to be reconstructable and trigger the measurement. We assume that either one of the existing CMS triggers (e.g. cosmic or dimuon trigger) would record these events, or that a dedicated, highly-efficient trigger can be designed, potentially exploiting the fact that those muon tracks would be close to parallel to the beam direction, as well as the timing of the collisions (the arrival time of muons would be possible to estimate and distinct from standard collisions at the CMS interaction point). The latter could also be exploited to suppress any potential beam-halo or cosmic background.

\begin{figure}[t!]
\centering
\includegraphics[width=.99\textwidth, trim={0 0 0 0}, clip]{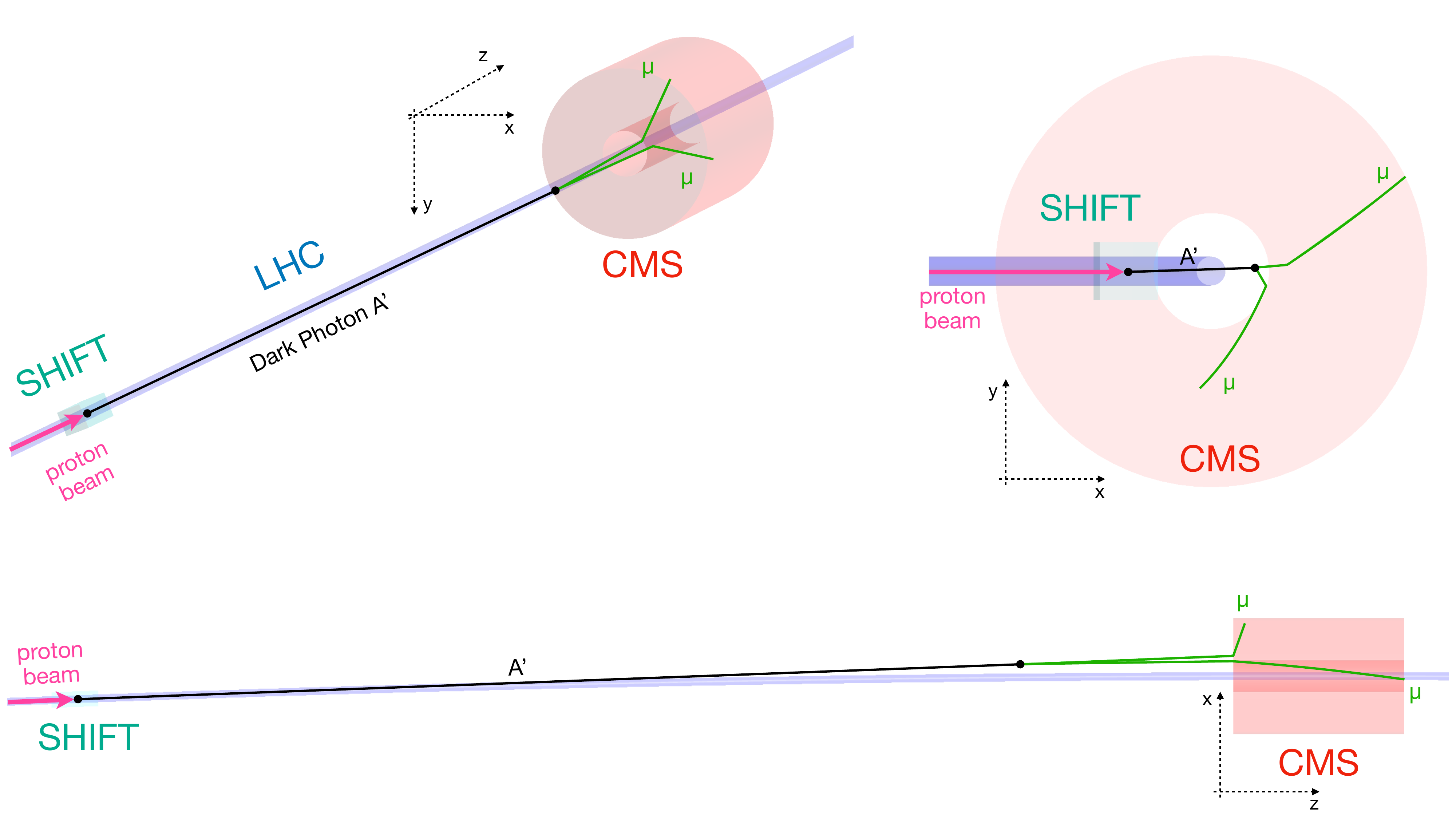}
\caption{\label{fig:event_display} Example visualization of a SHIFT event in a 3D view and 2D projections. The~event is a Dark Photon signal event with $m_{A'}=30$~GeV and a mean proper decay length of 10~m. The magnetic field of 2T is assumed, starting sharply at the edge of the detector.}
\end{figure}

The reconstruction of such horizontal tracks is currently difficult since the default algorithms are optimized for muons coming from the CMS interaction point (or at least from the general direction of the detector's center). For particles coming from SHIFT, the reconstruction algorithms would need to be tuned or a dedicated algorithm would have to be developed. Although this could be quite challenging because the solenoid magnet of CMS is designed to bend trajectories of particles with relatively large transverse momenta, it should be possible to achieve if the particles have enough energy to follow predictable trajectories. Muons coming from SHIFT are expected to cross at least 2-3 end-cap layers (usually more than that) and a few barrel layers, which is enough to reconstruct a track. An~example simplified event display is shown in Fig.~\ref{fig:event_display}, with the magnetic field of 2T starting sharply at the edge of the CMS detector. In this example, a Dark Photon decays around 20~meters before CMS into a collimated pair of muons, which get deflected upon reaching the detector. One can observe a significant imbalance in the longitudinal momentum of the two muons, resulting from the alignment or anti-alignment with the Dark Photon's longitudinal momentum in the mother's rest frame. This results in one of the muons traversing all of the CMS detector (potentially leaving many hits in the muon tracker) and the other one escaping quickly through the side. However, in most cases the imbalance is smaller than in this example, resulting in both muons leaving CMS through its sides after passing through roughly half of the detector. The details of the reconstruction need to be further studied but despite its challenges, there is no particular reason why such an algorithm, or a tune of the existing ones, could not be developed.

The muon tracks entering CMS through its end-caps have to trigger a measurement. Given the path of $\approx$150 meters and the products of the collision on the target traveling slower than the LHC beam, muons of interest will reach the detector many bunch crossings after the nominal collision. To distinguish between central and shifted collisions, the trigger should exploit the pointing direction of the muon track. Additional requirements on the dimuon vertex or the timing of the hits in the muons system may be imposed to reduce the trigger rate if needed. Finally, one can tune the position of the target to avoid signal muons arriving at the detector right between two bunch crossings.

Overall, the only requirement related to triggering and reconstruction is that muons have energy above 30 GeV. To be consistent between the CMS and SHIFT scenarios, we will apply the same criterion in both cases (instead of the usual transverse momentum requirement of CMS). Finally, we will assume that the trigger and reconstruction are 100\% efficient.

\subsection{Invariant mass range}

In order to simplify the study, a selection of $m^{\mu\mu} \in [11, 60]$ GeV is applied, rejecting the region populated by SM mesons decaying to pairs of muons, such as J/$\Psi$ or $\Upsilon$. \mbox{The~1--11~GeV} region is also interesting but would require a dedicated study, carefully taking into account SM resonances, invariant mass resolution, and other experimental effects. This is also a region in which many other experiments, such as aforementioned FASER, SND, or MATHUSLA, display sensitivity to various BSM models. A comparison with those experiments in this mass range could be performed using existing tools, such as \verb|SensCalc|~\cite{Ovchynnikov_2023}. At the high-mass end, a selection requirement is imposed to avoid interference with the SM Z boson.

\section{Results}

In this section, the results of the study will be presented, including efficiencies of selections for CMS and SHIFT scenarios, for backgrounds and different signal hypotheses, but also mass distributions of passing events, and expected cross-section limits. Here I would like to note that a scenario with a fixed target at LHCb was also considered - in general the acceptance (and therefore also the overall signal efficiency) for studied signals was found to be between that of CMS and SHIFT. However, with around 25 times lower total integrated luminosity than that of CMS, LHCb was found to have worse reach than CMS for almost all signal benchmark points (except 15 GeV Dark Photons with a mean proper decay length of 10 meters). It has never reached sensitivity similar to SHIFT, and therefore the details of this study are omitted in this paper. As explained in Sec.~\ref{sec:intro}, experiments such as FASER or MATHUSLA, due to their acceptance, using the collider mode, and other characteristics, target particles with masses up to $\approx$1~GeV (an order of magnitude below the focus of this study), and therefore will not be further discussed.

\subsection{Selection efficiency}
\label{sec:selection_efficiency}

The relative and total efficiencies of different selections, as well as total expected yields, are listed in Tab.~\ref{tab:cut_flow_backgrounds} for the background processes. First, one can notice that SHIFT is more efficient than CMS for the backgrounds. This may seem counter-intuitive at first, but can be understood when compared to kinematic distributions in Fig.~\ref{fig:initial_kinematics}. First, one should realize that the location of the SHIFT interaction point was tuned to maximize acceptance, given the $\eta^{\mu}$ distribution - this causes most muons produced in the collision to travel roughly in the direction of the CMS detector and therefore results in a high angular acceptance for DY processes (even higher than in CMS, where many muons fall outside of the covered $\eta$ range). Then, for the QCD background, the requirement of $E^{\mu} > 30$~GeV means that most of the muons travel in the forward direction (even in the collider mode), resulting in the acceptance requirement having three orders of magnitude higher impact on QCD in the CMS scenario than in the SHIFT one. Overall, DY processes are suppressed roughly by 1-2 orders of magnitude, while QCD is suppressed by 9-10 orders of magnitude.

\begin{table}[ht]
\centering
\resizebox{0.6\linewidth}{!}{
    \begin{tabular}{l|cc|cc}
    Selection                           & \multicolumn{2}{c|}{DY}               & \multicolumn{2}{c}{QCD}              \\
    \hline
                                        & CMS               & SHIFT             & CMS               & SHIFT             \\
    \hline
    $\ge2\mu$ with $E^{\mu} > 30$ GeV   & 0.5               & 0.8               & $1\cdot10^{-3}$   & 1.0               \\
    Within acceptance                   & 0.2               & 0.6               & $1\cdot10^{-6}$   & $1\cdot10^{-2}$   \\
    Material survival                   & 1.0               & 0.9               & 1.0               & $7\cdot10^{-3}$   \\
    $m^{\mu\mu} \in [11, 60]$ GeV       & $9\cdot10^{-2}$   & 0.7              & $3\cdot10^{-3}$   & $1\cdot10^{-5}$   \\
    \hline
    Total Efficiency                    & $8\cdot10^{-3}$   & $3\cdot10^{-1}$   & $4\cdot10^{-10}$  & $2\cdot10^{-9}$   \\
    Expected yield                      & $4\cdot10^{7}$    & $7\cdot10^{4}$    & $6\cdot10^{8}$    & $8\cdot10^{5}$    \\
    \end{tabular}
}
\caption{Efficiencies of each selection and expected yields for DY and QCD background processes in the CMS and SHIFT variants. For the yield calculation, a total integrated luminosity of $715~\rm{fb}^{-1}$ ($7.15~\rm{fb}^{-1}$) is used for the CMS (SHIFT) scenario.}
\label{tab:cut_flow_backgrounds}
\end{table}

The results for Dark Photons and Hidden Valley signals are listed in Tab.~\ref{tab:cut_flow_DP} and Tab.~\ref{tab:cut_flow_HV}, respectively. The selections are 1-2 orders of magnitude more efficient at SHIFT than at CMS for the Dark Photons scenario, and around 10 times more efficient for the Hidden Valley scenarios. As expected, the acceptance criteria are mainly affected by the proper decay length, while the $m^{\mu\mu} \in [11, 60]$~GeV requirement starts to play a role when going to lower DP or DH masses. Overall, the short-lived signal efficiency is slightly lower than background efficiency at CMS, but two times higher at SHIFT.

\begin{table}[ht]
\centering
\resizebox{\linewidth}{!}{
    \begin{tabular}{l|cc|cc|cc|cc}
                                        & \multicolumn{6}{c|}{$m_{DP}=30~\rm{GeV}$}                                                                             & \multicolumn{2}{c}{$m_{DP}=5~\rm{GeV}$} \\
    Selection              & \multicolumn{2}{c}{$c\tau = 10^{-3}~\rm{m}$} & \multicolumn{2}{c}{$c\tau = 10^{1}~\rm{m}$} & \multicolumn{2}{c|}{$c\tau=10^{3}~\rm{m}$} & \multicolumn{2}{c}{$c\tau = 10^{1}~\rm{m}$} \\
    \hline
                                        & CMS               & SHIFT             & CMS               & SHIFT             & CMS               & SHIFT             & CMS               & SHIFT             \\
    \hline
    $\ge2\mu$ with $E^{\mu} > 30$ GeV   & 0.5               & 0.9               & 0.5               & 0.9               & 0.5               & 0.9               & 0.4               & 0.8               \\
    Within acceptance                   & 0.1               & 0.6               & $6\cdot10^{-3}$   & 0.2               & $7\cdot10^{-5}$   & $3\cdot10^{-3}$   & $2\cdot10^{-3}$   & 0.2               \\
    Material survival                   & 1.0               & 1.0               & 1.0               & 1.0               & 1.0               & 1.0               & 1.0               & 1.0               \\
    $m^{\mu\mu} \in [11,60]$ GeV        & 1.0               & 1.0               & 1.0               & 1.0               & 0.9               & 0.9               & 0.4               & 0.6               \\
    \hline
    Total Efficiency                    & $5\cdot10^{-2}$   & $5\cdot10^{-1}$   & $3\cdot10^{-3}$   & $2\cdot10^{-1}$   & $3\cdot10^{-5}$   & $2\cdot10^{-3}$   & $4\cdot10^{-4}$   & $9\cdot10^{-2}$   \\
    \end{tabular}
}
\caption{Efficiencies of each selection for different Dark Photon scenarios.}
\label{tab:cut_flow_DP}
\end{table}


\begin{table}[ht]
\centering
\resizebox{0.6\linewidth}{!}{
    \begin{tabular}{l|cc|cc}
    Selection                           & \multicolumn{2}{c|}{$m_{DH}=20$~GeV}  & \multicolumn{2}{c}{$m_{DH}=5$~GeV}   \\
    \hline
                                        & CMS               & SHIFT             & CMS               & SHIFT             \\
    \hline
    $\ge2\mu$ with $E^{\mu} > 30$ GeV   & 0.7               & 1.0               & 0.6               & 1.0               \\
    Within acceptance                   & 0.3               & 1.0               & 0.1               & 1.0               \\
    Material survival                   & 1.0               & 1.0               & 1.0               & 1.0               \\
    $m^{\mu\mu} \in [11, 60]$ GeV               & 1.0               & 1.0               & 0.5               & 0.7               \\
    \hline
    Total Efficiency                    & $2\cdot10^{-1}$   & 1.0   & $5\cdot10^{-2}$   & $7\cdot10^{-1}$   \\
    \end{tabular}
}
\caption{Efficiencies of each selection for Hidden Valley scenarios with $m_{Z'}=60$~GeV, $c\tau=10^{-1}$~m and different Dark Hadron masses.}
\label{tab:cut_flow_HV}
\end{table}

\begin{figure}[t!]
\centering
\includegraphics[width=.49\textwidth, trim={0 0 70pt 0}, clip]{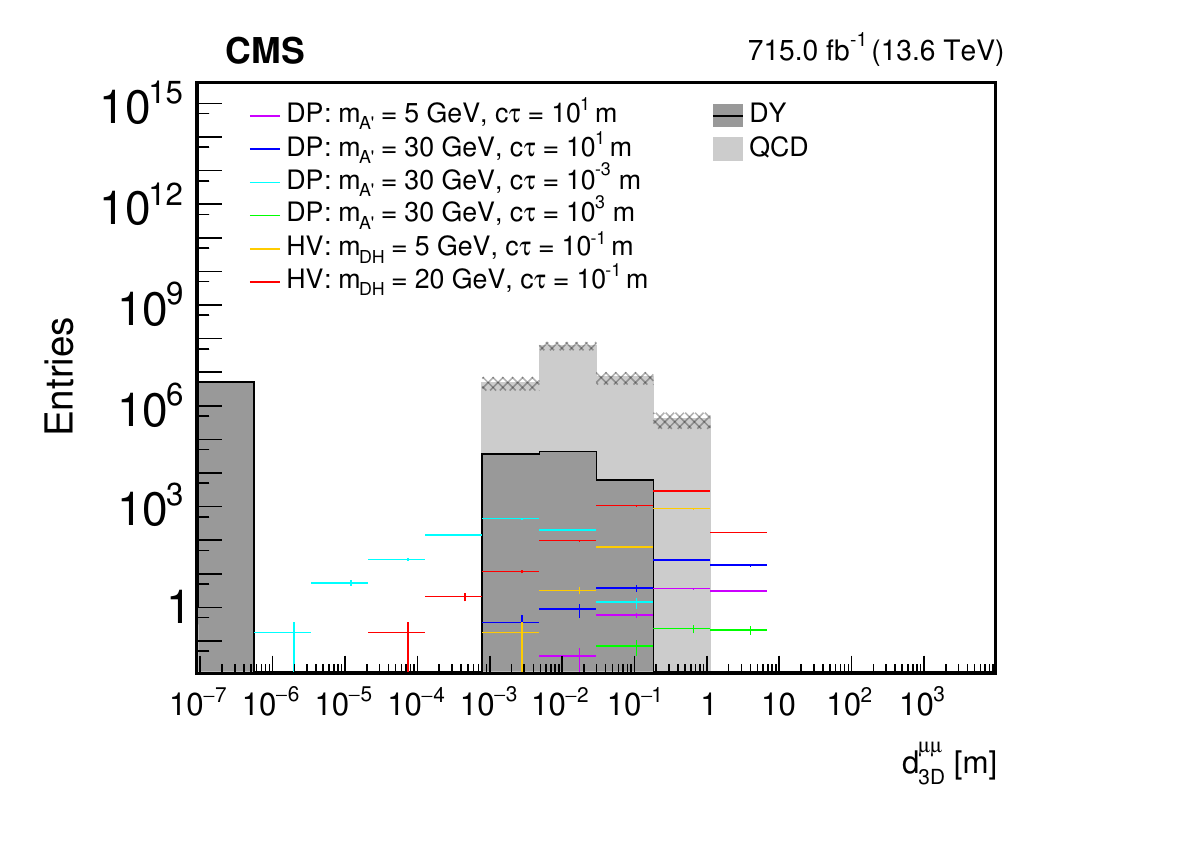}
\includegraphics[width=.49\textwidth, trim={0 0 70pt 0}, clip]{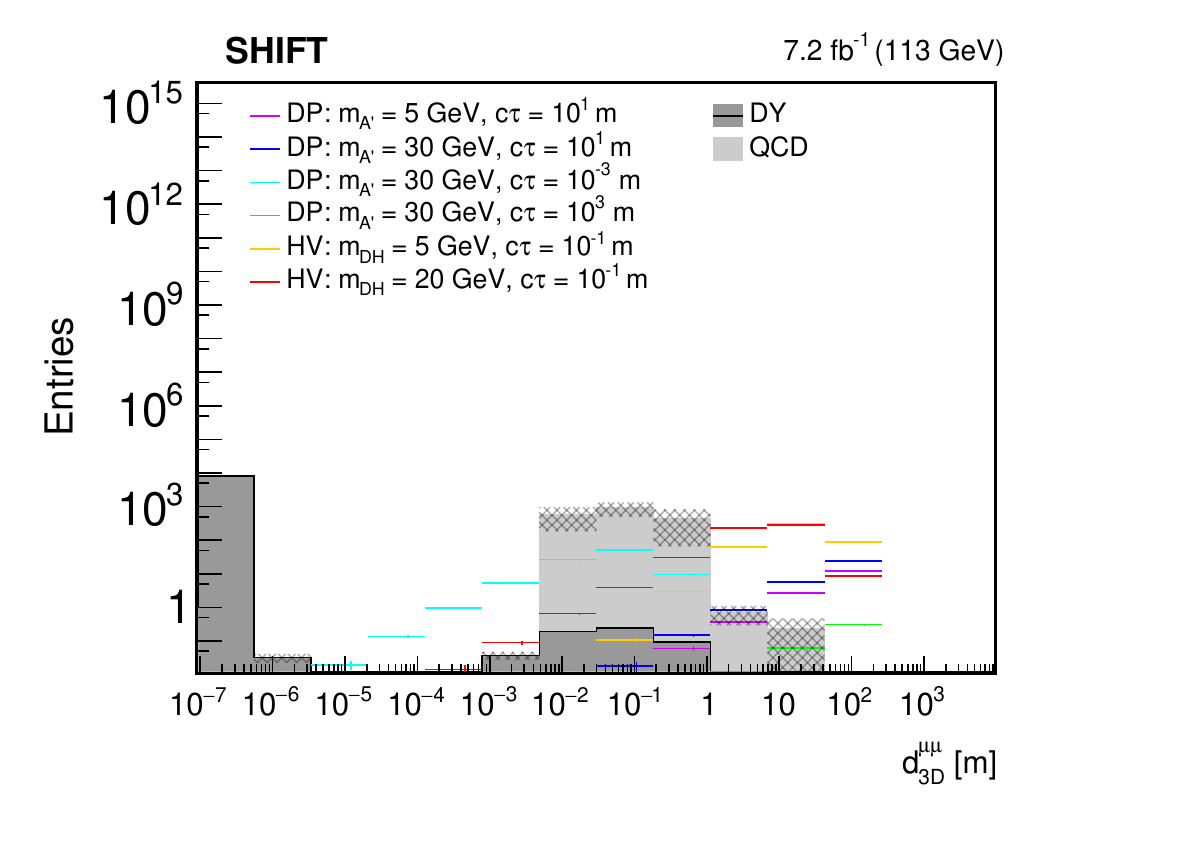}
\includegraphics[width=.49\textwidth, trim={0 0 70pt 0}, clip]{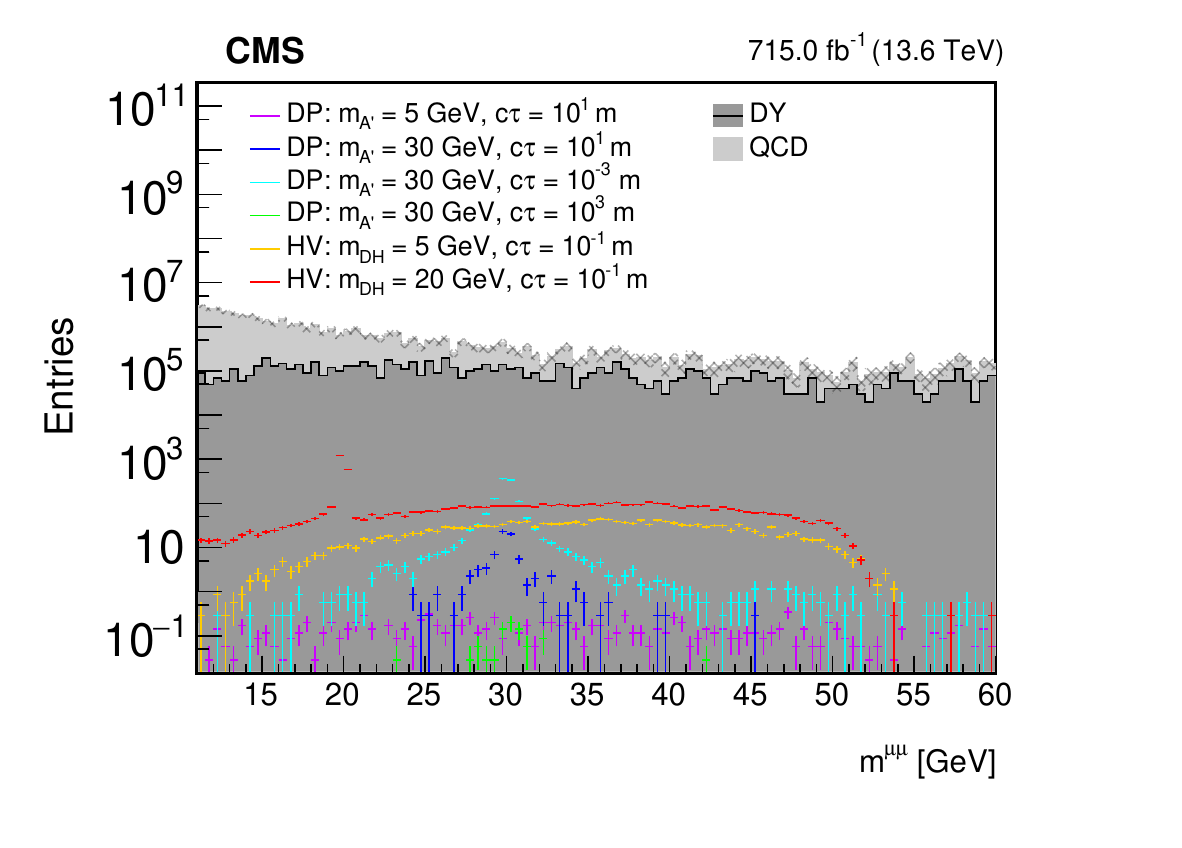}
\includegraphics[width=.49\textwidth, trim={0 0 70pt 0}, clip]{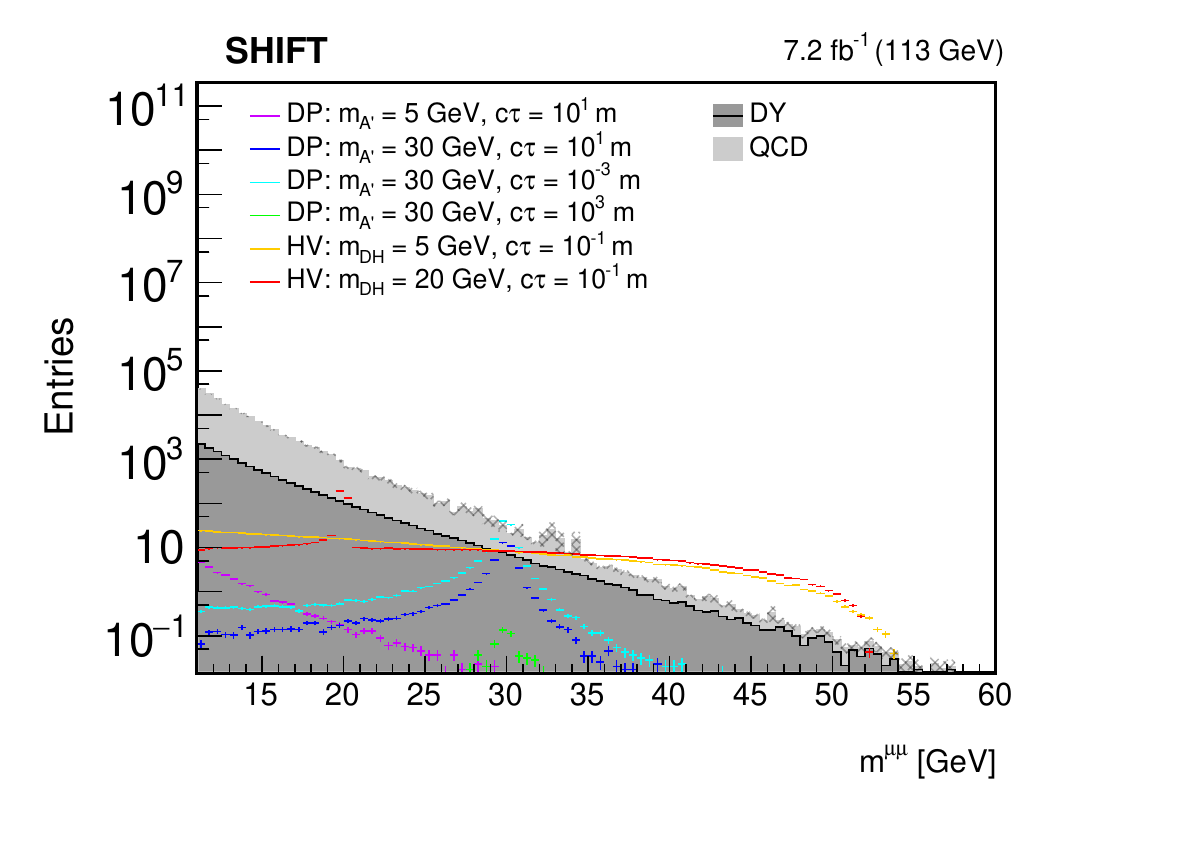}
\caption{\label{fig:final_pair_kinematics} Distance of the dimuon vertex from the collision point (top) and dimuon invariant mass (bottom), after passing all selections. Left: CMS scenario, right: SHIFT scenario. Distributions are normalized to cross-section and luminosity ($715~\rm{fb}^{-1}$ for CMS and $7.15~\rm{fb}^{-1}$ for SHIFT), with cross-section set for all signals to $0.2~\rm{pb}$~(nb) for SHIFT (CMS) for visual purposes.}
\end{figure}

The distributions of dimuon invariant masses and the dimuon vertex distance from the collision point $d^{\mu\mu}_{3D}$, for backgrounds and different signal hypotheses after passing all selections, are shown in Fig.~\ref{fig:final_pair_kinematics}. The binning of the mass distribution for limits setting is chosen such that bin width is 0.5 GeV, which is conservatively larger than the CMS dimuon mass resolution (at the level of 1-2\% at 10~GeV). All combinations of muons enter the dimuon distributions, without any requirements on both of them coming from the same vertex. However, it was verified that placing requirements on a common vertex has a~negligible effect on the limits presented below.

A few interesting features can be observed. First, the difference in lifetime acceptance is visible in the $d_{3D}^{\mu\mu}$ distribution being restricted to be within a few meters in CMS and extending to above 100 meters in SHIFT. Then, in the $m^{\mu\mu}$ distribution one can see that Dark Photons with the mass of 5 GeV (below the dimuon mass requirement) are indistinguishable from the DY background - as a result, searching for them would be very difficult both at CMS and SHIFT. However, the Dark Photons at 30 GeV display a clear peak, which can be exploited in a search for such particles. The magnitude of said peaks decreases with the increasing proper decay length (due to lower selection efficiency). As a result of the different underlying physics of the Dark Sector, the Dark Hadron spectra have a shape significantly different from the SM background and the DP signal (the latter being tuned to be similar to the SM). In this case, even the tail of the 5 GeV signal could be used for a search, despite the peak being below the allowed mass range. Nevertheless, moving to higher masses (e.g.~20~GeV) allows one to exploit the peak of the Dark Hadron (on top of the different overall shape of the distribution), resulting in stronger limits.

\subsection{Limits}
\label{sec:limits}

In this section, expected limits on different BSM scenarios are presented. In all cases, the dimuon invariant mass distributions (see Fig.~\ref{fig:final_pair_kinematics}) are used to calculate asymptotic limits at 95\% confidence level, using the CMS \verb|combine| tool \cite{cmscollaboration2024cms}. A systematic uncertainty of 1.5\% on the integrated luminosity is assumed \cite{CMS-PAS-LUM-22-001}. In addition, a flat 10\% systematic uncertainty is included to partially account for some of the simplifications and experimental effects that were not studied in detail in this work.

\begin{figure}[ht]
\centering
\includegraphics[width=.53\textwidth]{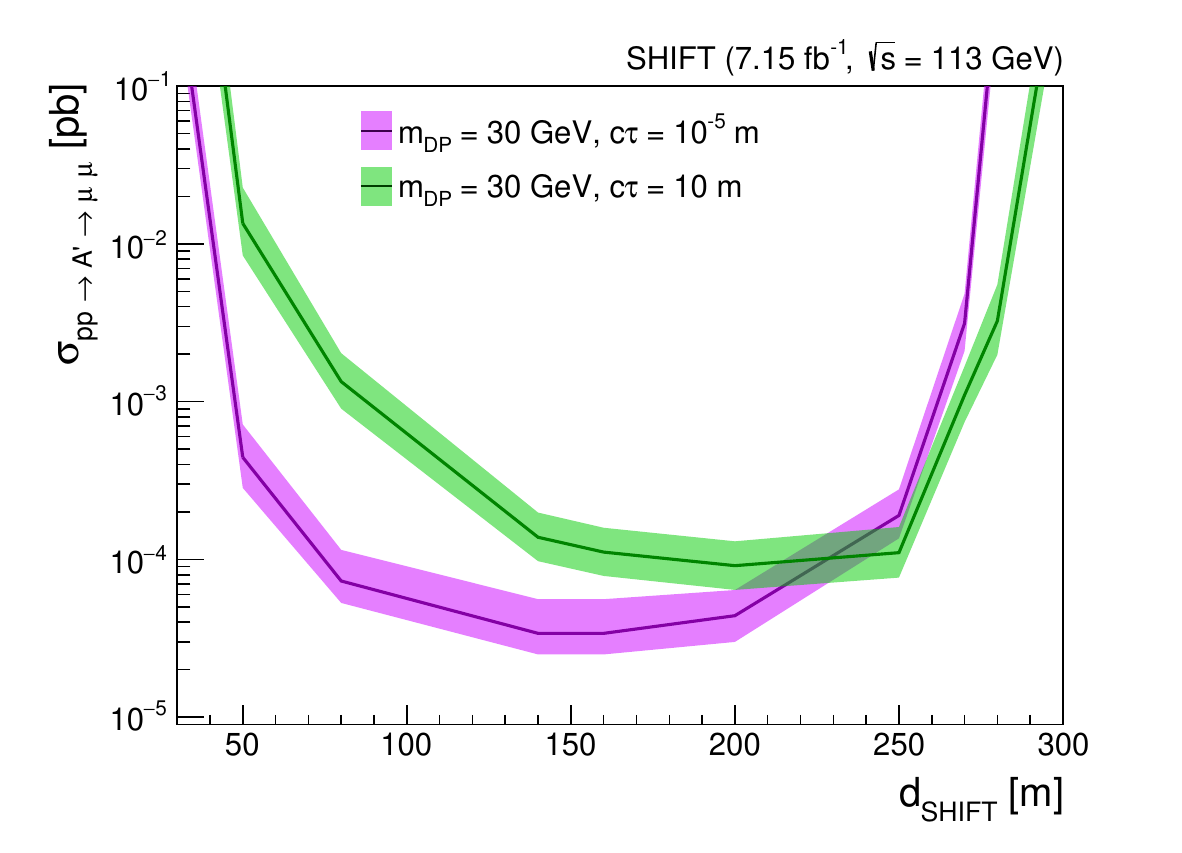}
\caption{\label{fig:limits_DP_distance} Expected 95\% CL limits from SHIFT@LHC on production cross-section for Dark Photons with $m_{DP}=30$~GeV and two different lifetimes, as a function of the distance between SHIFT fix target and the center of the CMS detector. The lines show the central value, while the bands are the 1$\sigma$ uncertainty ranges.}
\end{figure}

The limits on the cross-section for Dark Photons as a function of the distance between the SHIFT fixed target and the center of the CMS detector are shown in Fig.~\ref{fig:limits_DP_distance}. As can be seen, depending on the signal hypotheses, distances between 150 and 250 meters provide the best coverage. Based on the optimization performed on a larger number of signal samples, a distance of 160~meters was fixed for all other results presented in this paper.

\begin{figure}[ht]
\centering
\includegraphics[width=.49\textwidth]{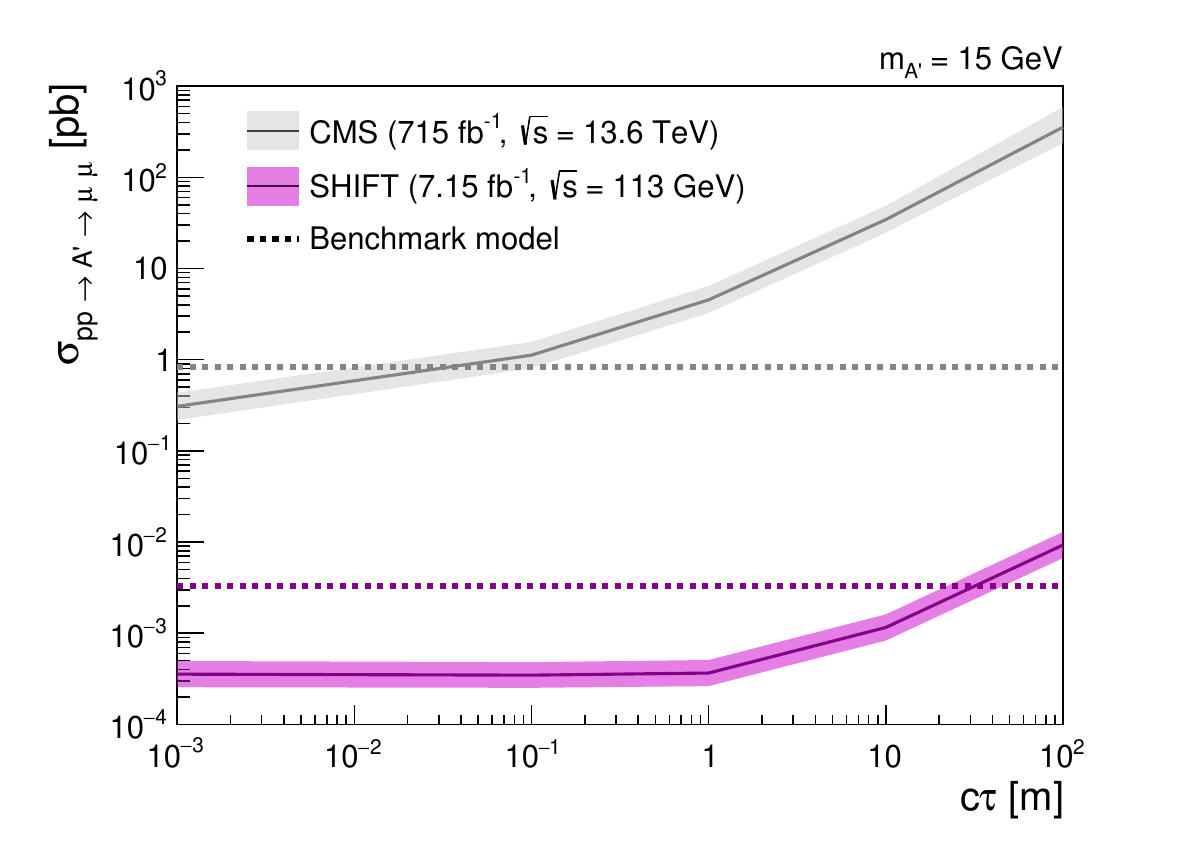}
\includegraphics[width=.49\textwidth]{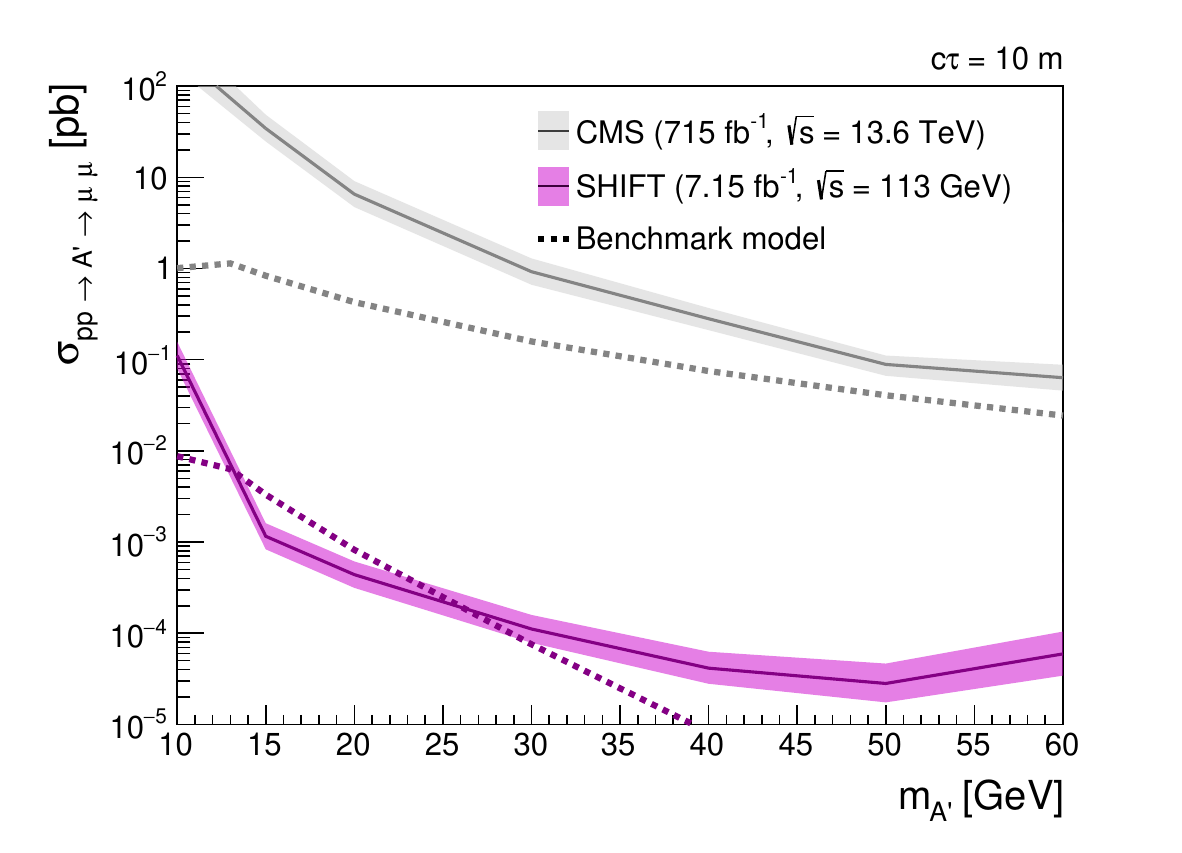}
\caption{\label{fig:limits_DP_ctau_mDarkPhoton} Comparison of SHIFT@LHC (z=160 m) and CMS 95\% CL limits on production cross-section for Dark Photons with a mass of 15 GeV as a function of proper decay length (left) and Dark Photons with the proper decay length of 10 m as a function of mass (right). The lines show the central value, while the bands are the 1$\sigma$ uncertainty ranges. The luminosity is $715~\rm{fb}^{-1}$ for CMS and $7.15~\rm{fb}^{-1}$ for SHIFT.}
\end{figure}

With a fixed distance of 160~meters, Fig.~\ref{fig:limits_DP_ctau_mDarkPhoton} presents limits on the cross-section for Dark Photons as a function of the mean proper decay length and the Dark Photon mass. The~limits alone are much more stringent for SHIFT than CMS, however, the production cross-section is also expected to be much smaller at $\sqrt{s}=$113 GeV compared to 13.6 TeV. For~this reason, limits are compared to benchmark cross-sections calculated at different center of mass energies. As can be seen, with only 1\% of the CMS luminosity expected for Run~4, SHIFT can exclude mean proper lifetimes up to 30~meters, with CMS only reaching 4~cm. With the mean proper lifetime of 10~meters CMS would not reach the benchmark model line, while SHIFT would cover the region from $\approx$14~GeV to 25 GeV.

\FloatBarrier

\begin{figure}[ht]
\centering
\includegraphics[width=.56\textwidth]{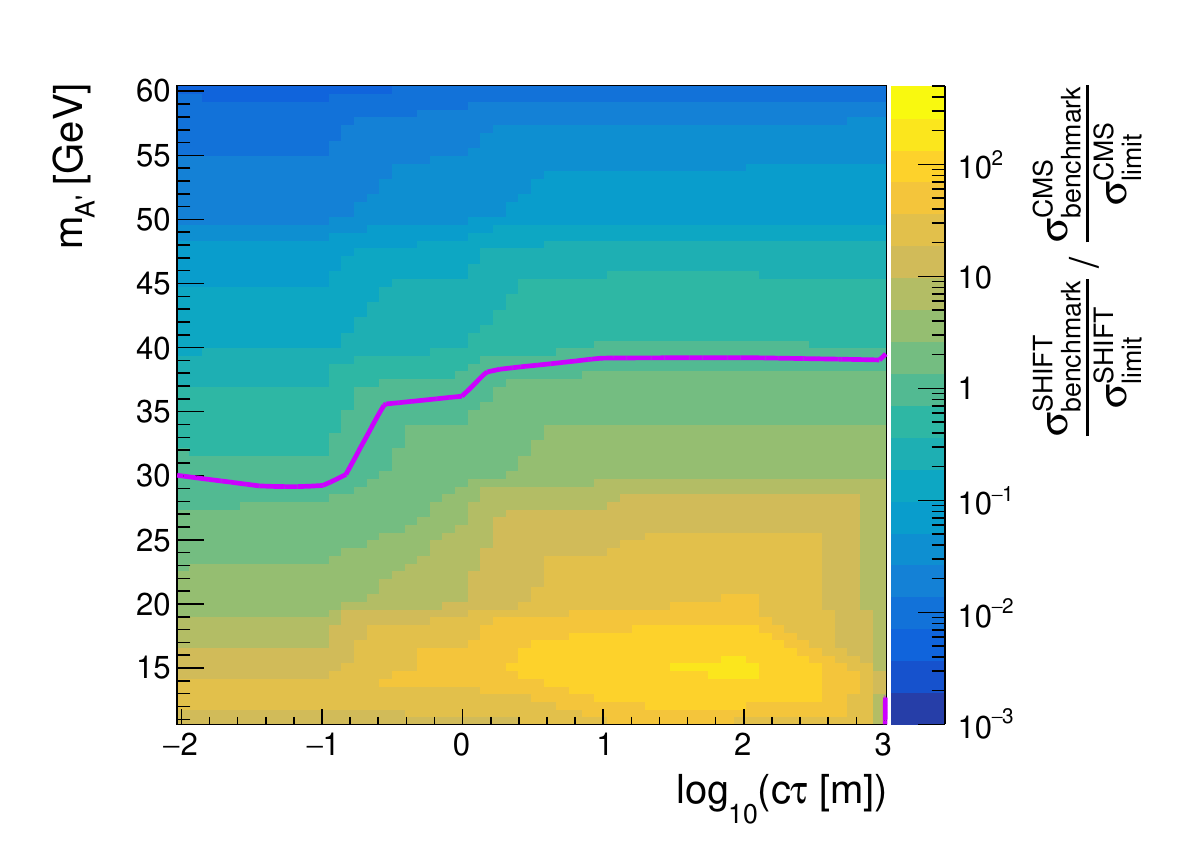}
\caption{\label{fig:limits_DP_2d} Ratio of the exclusion strength of SHIFT: $\sigma^{SHIFT}_{benchmark} / \sigma^{SHIFT}_{limit}$ and the exclusion strength of CMS: $\sigma^{CMS}_{benchmark} / \sigma^{CMS}_{limit}$, using 95\% CL limits on the production cross-section for Dark Photons compared to the benchmark cross-section, as a function of proper decay length and DP mass. The luminosity is $715~\rm{fb}^{-1}$ for CMS and $7.15~\rm{fb}^{-1}$ for SHIFT. The region enclosed by the purple contour shows where SHIFT has a better reach than CMS.}
\end{figure}

Figure~\ref{fig:limits_DP_2d}~shows Dark Photon limits in a 2D plane with the mean proper decay length on the x-axis, the DP mass on the y-axis, and a SHIFT over CMS double ratio on the z-axis. This double ratio is calculated as the exclusion strength of SHIFT: $\sigma^{SHIFT}_{benchmark} / \sigma^{SHIFT}_{limit}$, divided by the exclusion strength of CMS: $\sigma^{CMS}_{benchmark} / \sigma^{CMS}_{limit}$. The purple contour shows where the double ratio equals 1.0, therefore enclosing a region in which SHIFT has an advantage over CMS. As expected, SHIFT provides better limits at low masses and high lifetimes, with over a factor 150 improvement at $\rm{m}_{DP}\approx15$~GeV and c${\tau}\approx100$~meters.

\FloatBarrier

\begin{figure}[t!]
\centering
\includegraphics[width=.49\textwidth]{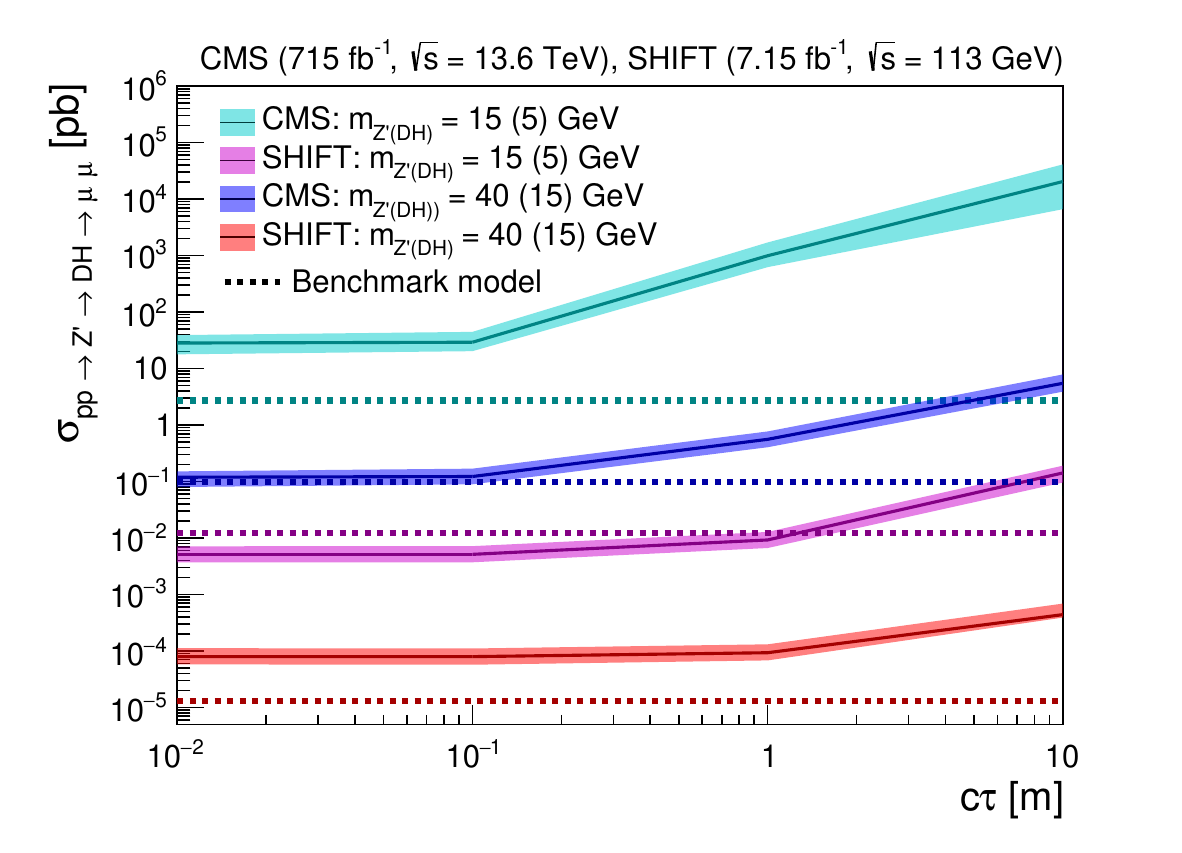}
\includegraphics[width=.49\textwidth]{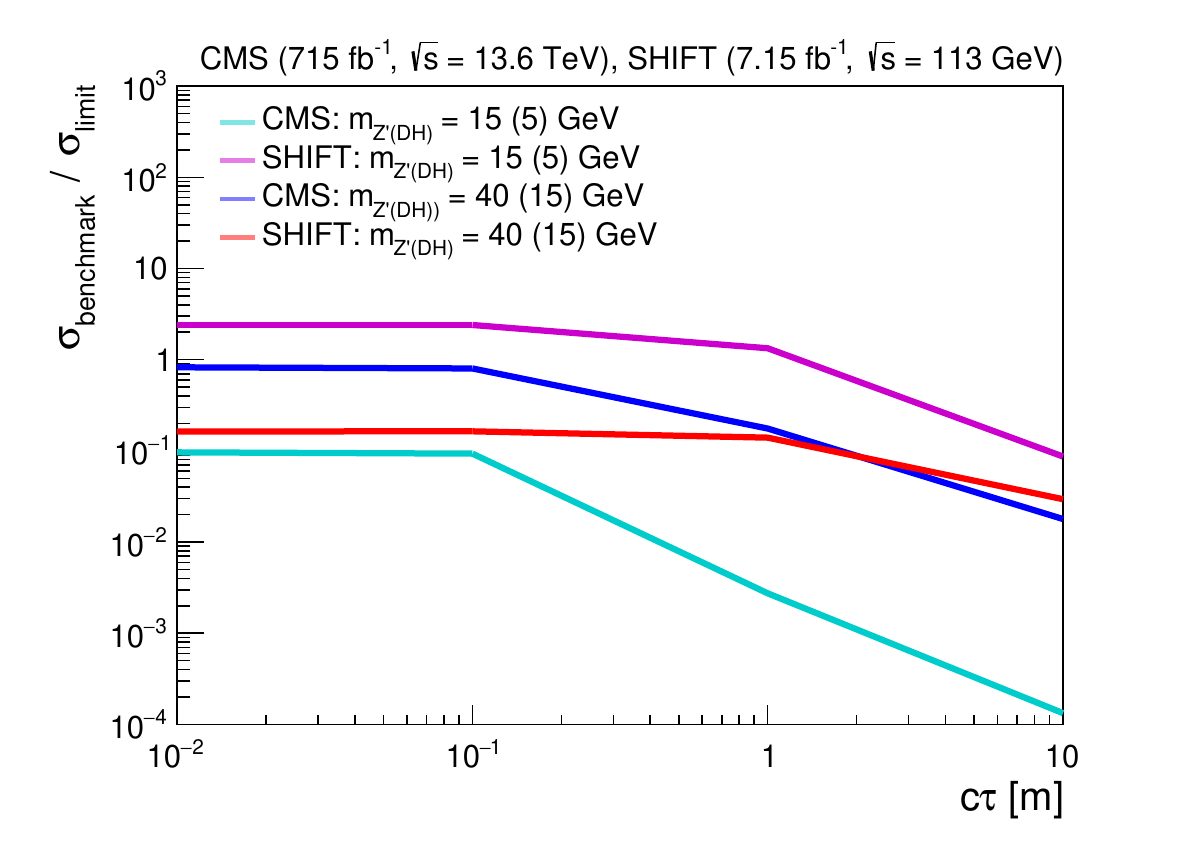}
\caption{\label{fig:limits_HV} Comparison of SHIFT@LHC (z=160 m) and CMS 95\% CL limits on production cross-section for two Hidden Valley scenarios (cyan/magenta: m$_{Z'} = 15$~GeV, m$_{DH}$ = 5~GeV and blue/red: m$_{Z'} = 40$~GeV, m$_{DH}$ = 15~GeV) as a function of mean proper lifetime. For~both scenarios m$_{DQ}$ = 1 GeV. Left: the solid lines show the central value, the bands are the 1$\sigma$ uncertainty ranges, and the dashed lines show the benchmark cross-sections. Right: ratios of benchmark cross-section over the limit in cross-section for different scenarios.}
\end{figure}

Limits for the Hidden Valley model are presented in Fig.~\ref{fig:limits_HV} for two scenarios: low-mass with $m_{Z'} = 15$~GeV, $m_{DH}=5$~GeV, for which only the different slope of the dimuon invariant mass distribution can be used, and mid-mass with $m_{Z'}=40$~GeV, $m_{DP}=15$~GeV, where also the peak can be exploited. As expected, limits for the mid-mass are in general stronger than those for the low-mass. In the low-mass case, SHIFT provides a factor of 10 improvement over CMS for prompt signatures, and up to 3 orders of magnitude improvement at a $c\tau$ of 10 meters, while CMS performs better than SHIFT in the mid-mass scenario, with the two getting closer as the mean proper lifetime increases.

\clearpage
\newpage

\section{Summary and outlook}
\label{sec:summary}

In this work, I propose to install a gaseous fixed target at the LHC (SHIFT), located around 160~meters downstream of the CMS collision point, and using the CMS detector to register decay products (e.g. muons) of new particles with masses accessible at the lower center of mass energies of $\approx$113~GeV. The impact of different event kinematics, the angular and lifetime acceptance, as well as the survival probability of muons, was studied. The physics potential of such a program is assessed using two BSM hypotheses: a Dark Photon model and a Hidden Valley model, and the results are presented for SHIFT@LHC compared to the standard proton-proton CMS program in the collider mode.

This first study makes several assumptions and simplifications, especially regarding the fixed target material and exact location, the available luminosity, and lacks the precise simulation of the rock, magnets, and other material located between the fixed target and the detector. I acknowledge that a much more detailed study of these aspects would be needed to estimate physics potential more precisely. Nevertheless, the results of this preliminary study show that despite assuming just 1\% of the CMS luminosity in Run 4, the physics reach for both Dark Photons and Hidden Valley models can be improved by well over two orders of magnitude, depending on the model parameters. What is worth emphasizing here is that the feasibility of installing such a fixed target at the LHC has been already demonstrated by SMOG and the LHCb Collaboration. This solution is also relatively inexpensive, compared to building a new dedicated detector or even a dedicated new cavern, as is the case for many other proposed and existing extensions of the LHC physics program.

While already very promising, one should realize that these two studied models are just the tip of an iceberg: implementation of SHIFT would open up a vast space for searches, not only in scenarios with muons in the final state, but also electron, photons, jets, and hadrons. What is also interesting to consider is that a fixed target would naturally contain not only protons but also electrons, which would allow for quark-electron collisions, greatly amplifying cross-sections for direct production of leptoquarks. A large number of BSM models could be studied with SHIFT, giving access to otherwise uncovered corners of the parameter phase space. 

\acknowledgments

I would like to thank Juliette Alimena and Freya Blekman for their support, thoughtful discussions of the ideas presented in this work, and thorough review of the manuscript. I also acknowledge the support from DESY (Hamburg, Germany), a member of the Helmholtz Association HGF, and support by the Deutsche Forschungsgemeinschaft (DFG, German Research Foundation) under Germany’s Excellence Strategy -- EXC 2121 "Quantum Universe" -- 390833306.


\newpage

\bibliographystyle{JHEP}
\bibliography{shift}

\end{document}